\documentclass[twocolumn]{aastex62}
\usepackage[latin1]{inputenc}
\usepackage[english]{babel}
\usepackage{amsmath}
\usepackage{amsfonts}
\usepackage{amssymb}
\usepackage{makeidx}
\usepackage{graphicx}
\usepackage{color}
\usepackage{appendix}
\usepackage{multirow}
\usepackage[flushleft]{threeparttable}

\newcommand{\derd}{{\rm d}}
\newcommand{\Log}{{\rm Log}}
\newcommand{\Ms}{{\rm M}_\odot}

\newcommand{\gc}{{\rm SC}}
\newcommand{\SC}{{\rm SC}}

\newcommand{\bh}{{\rm SMBH}}
\newcommand{\ibh}{{\rm IMBH}}
\newcommand{\nc}{{\rm NC}}
\newcommand{\nb}{{\rm NB}}
\newcommand{\df}{{\rm DF}}

\graphicspath{{./}{./}}

\submitjournal{ApJL}

\shorttitle{The origin of a rotating metal-poor stellar population in the MWNC}
\shortauthors{Arca Sedda, M. et al}

\begin{document}

\title{On the origin of a rotating metal-poor stellar population in the Milky Way Nuclear Cluster}

\correspondingauthor{Manuel Arca Sedda}
\email{m.arcasedda@gmail.com}

\author[0000-0002-3987-0519]{Manuel Arca Sedda}
\affil{Astronomisches Rechen-Institut, Zentrum f\"{u}r Astronomie der Universit\"{a}t  Heidelberg, M\"onchhofstr. 12-14, D-69120 Heidelberg, Germany}

\author[0000-0002-9420-2679]{Alessia Gualandris}
\affil{Department of Physics, University of Surrey, Guildford, GU2 7XH, United Kingdom}

\author[0000-0001-9554-6062]{Tuan Do} 
\affil{UCLA Galactic Center Group, Physics and Astronomy Department, UCLA, Los Angeles, CA 90095-1547}

\author[0000-0002-0160-7221]{Anja Feldmeier-Krause}
\affiliation{The Department of Astronomy and Astrophysics, The University of Chicago, 5640 S. Ellis Ave, Chicago, IL 60637, USA}

\author[0000-0002-6922-2598]{Nadine Neumayer}
\affil{Max Planck-Instit\"ut f\"ur Astronomie, Konigstuhl 17, D-69117 Heidelberg, Germany}

\author[0000-0002-8448-5505]{Denis Erkal}
\affil{Department of Physics, University of Surrey, Guildford, GU2 7XH, United Kingdom}

\begin{abstract}
We explore the origin of a population of stars recently detected in the inner parsec of the Milky Way Nuclear Cluster (NC), which exhibit sub-solar metallicity and a higher rotation compared to the dominant population. Using state-of-the-art $N$-body simulations, we model the infall of a massive stellar system into the Galactic center, both of Galactic and extra-galactic origin. We show that the newly discovered population can either be the remnant of a massive star cluster formed a few kpc away from the Galactic center (Galactic scenario) or be accreted from a dwarf galaxy originally located at $10-100$ kpc (extragalactic scenario) and that reached the Galactic center $3-5$ Gyr ago. A comparison between our models and characteristic Galactocentric distance and metallicity distributions of Milky Way satellites and globular clusters favours the Galactic scenario. A comparison with clusters associated with the Enceladus-Sausage, Sequoia, Sagittarius and Canis Major structures suggests that the progenitor of the observed metal-poor substructure formed in-situ rather than being accreted.
\end{abstract}


\section{Introduction}

Like the majority of galaxies in the Universe, the center of the Milky Way (MW) harbours a supermassive black hole (SMBH), SgrA*, with a mass $M_\bh = (4.0-4.3) \times 10^6 \Ms$ \citep{ghez08,gillessen17}, surrounded by a dense and massive nuclear star cluster (NC) with a total mass of $M_\nc  = 2.5\times 10^7\Ms$ \citep{schodel14,feldmeier17,neumayer20} and effective radius $R_e = 3.8-5.1$ pc \citep{schodel14,gallegocano20}. The Galactic center is the closest galactic nucleus, thus representing a unique target to study the interplay between an SMBH and its environment.
In our companion paper (Do et al., in press), we present observations of a region of the Galactic center centered on SgrA* and with a projected radius of $2.6$ pc$\times1.6$ pc based on the dataset and metallicity measurements obtained with the KMOS spectrograph on the Very Large Telescope (VLT) \citep{feldmeier17a}. This seeing-limited spectroscopic survey consists of about 700 late-type giant stars with ages of about 3-10 Gyr. Their spectral-type, metallicity and radial velocities were measured via full-spectrum fitting \citep{kerzendorf15,feldmeier17a}.  
These observations revealed the presence of a sub-population of stars characterised by lower metallicity and higher rotation than the overall NC population, which may be the relics of a star cluster spiralled in via dynamical friction \citep{feldmeier14}, as suggested by numerical models \citep{tsatsi17}. 

In this letter, we use a high-resolution $N$-body simulation to explore whether these chemical and kinematical features can be explained with a recent infall of either a massive and dense stellar cluster or a dwarf galaxy nucleus into the Galactic center. Such an event would support the so-called ``dry-merger'' scenario for NCs, according to which NCs are built, at least partially, by the inspiral and merging of star clusters (SCs) via dynamical friction \citep{TrOsSp,Dolc93}. This scenario accounts for several observational features of both the Galactic \citep{AMB,perets14,tsatsi17} and extragalactic NCs \citep{antonini13,ASCD15He,ASCD17}, although it likely operates in concert with in-situ star formation \citep{NEUM11,antonini13,guillard16}, which can explain the complex star-formation histories observed in the majority of NCs \citep[see e.g.][]{rossa,seth06}. 
Alternatively, the low metallicity population might be the result of an old in-situ star formation episode following inflow of metal poor gas into the Galactic center. However, relaxation processes would likely have erased the peculiarities of the population \citep[e.g.][]{alexander05}.

This letter is organized as follows: we present the numerical method used to test the infall scenario in Section \ref{sec:met}, we show the results of our analysis in Section \ref{sec:res}, we discuss the implications of our simulations in Section \ref{sec:dis} and summarize our conclusions in Section \ref{sec:con}.

\section{Numerical method}
\label{sec:met}
To study the possible origin of the rotating metal-poor stellar population presented in our companion paper, we analyze data from one of the direct summation $N$-body simulations presented in \cite{ASG17}. Specifically, we use the model denominated ``Ma'', that simulates the orbital decay of a massive SC in a galactic nucleus composed of a nuclear bulge, a NC, and an SMBH. In the simulation, both the SC and the inner 150 pc of the galactic nucleus are modelled with a total of $N=1,048,576$ particles, corresponding to a mass resolution of $m_* \simeq 45\Ms$. This represents the current state-of-the-art for simulations of this type. The nuclear bulge is modelled as a truncated \cite{Plum} sphere with a total mass of $M_{\nb} = 3\times 10^{10}\Ms$ \citep{valenti16}, a length scale $r_{\nb} = 1$ kpc, and a truncation radius of $r_t = 150$ pc. This choice leads to a circular velocity profile compatible with observations of the Galactic bulge \citep{portail15}. The NC is modelled as a non-rotating \cite{Deh93} sphere with inner slope $\gamma \simeq 2$, scale radius $r_\nc = 4$ pc, and total mass $M_\nc = 10^7\Ms$. The central SMBH is accounted for as a point-like particle with mass $M_\bh = 5\times 10^6\Ms$. The SC is modelled as a \cite{King} sphere with total mass $M_\SC = 10^6 \Ms$, core radius $r_c = 0.24$ pc, and adimensional concentration parameter $W_0 = 6$ that initially moves on a circular orbit at a distance of $r = 50$ pc from the Galactic center. The center of the SC contains an intermediate mass black hole (IMBH) with mass $M_\ibh = 10^4\Ms$. The presence of an IMBH of such mass does not affect the infalling process nor the mass loss suffered by the SC as it migrates inward \citep[see Figures 3 and 7 in][]{ASG17}. Moreover, the IMBH is not expected to affect the relaxation process of the SC on the short ($< 100$ Myr) time-scale of the inspiral, nor the relaxation of the galactic nucleus, given the small IMBH to SMBH mass ratio.

We note that our SC model can also be interpreted as the remaining nucleus of a dwarf galaxy that merged with the MW and lost its stellar envelope, a scenario suggested to explain the properties of several globular clusters observed in the MW halo, including $\omega$Cen \citep{hilker00} and M54 \citep{alfaro19,alfaro20}. 
We stress that the numerical setup adopted here does not rely on any assumption on the earlier SC evolution. Our working hypothesis is that the SC progenitor formed outside the Galactic center and slowly migrated inward, losing stars on its way due to the Galactic tidal field and eventually merging with the NC. Our only requirement is that the SC reaches the inner 50 pc preserving a clear structure and a mass $\sim 10^6\Ms$. This is further discussed in Section \ref{sec:origin} where we place constraints on the SC history. 

\section{Results}
\label{sec:res}

\subsection{Star Cluster infall and merger with the Galactic Nuclear Cluster}
\label{sec:scinfall}

\begin{figure*}
\centering
\includegraphics[width=0.9\textwidth]{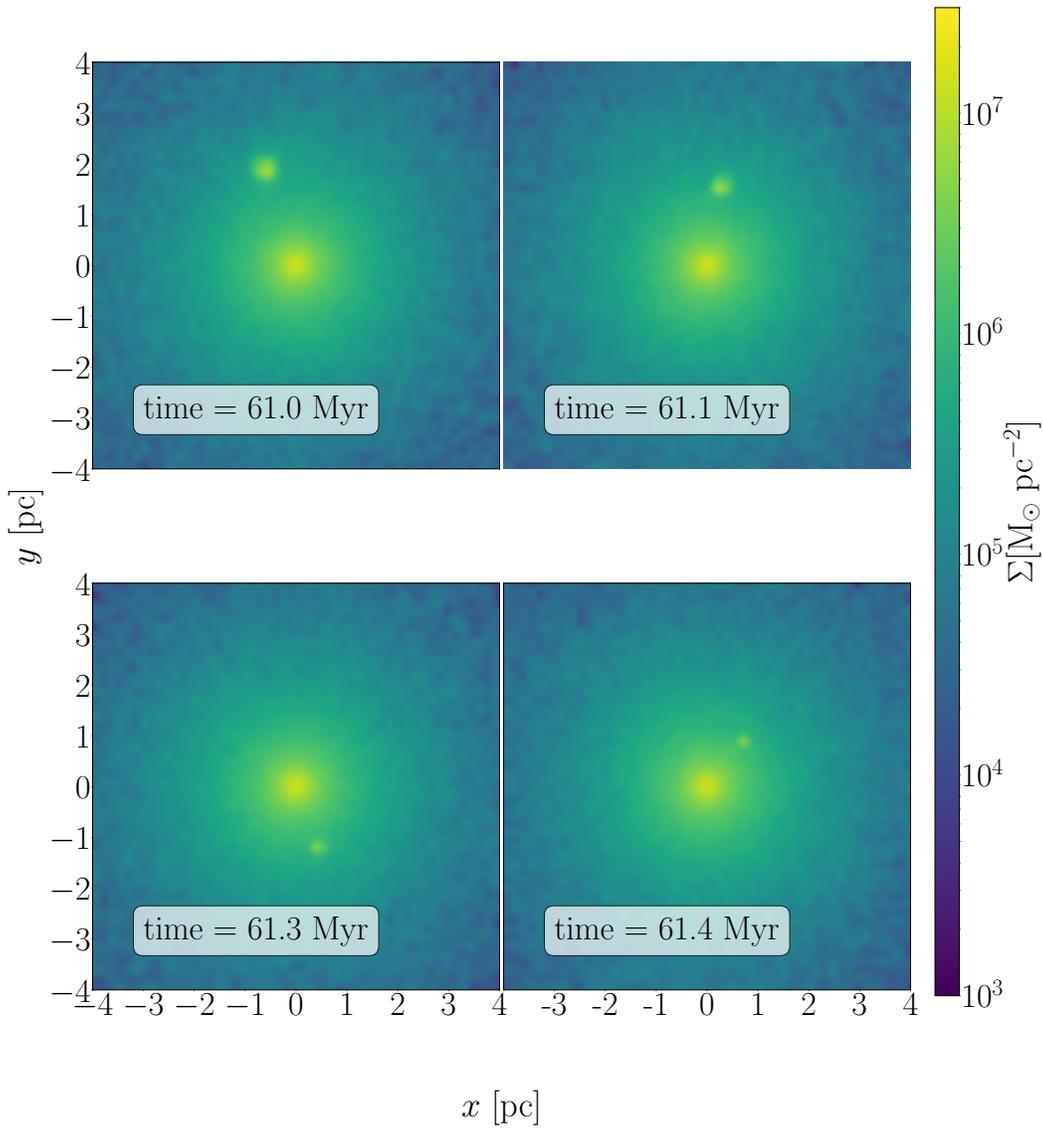}
\caption{Isodensity maps of the innermost 4 parsec region in the Galactic center, showing the NC overdensity and the infalling SC at different times. The SC rapid destruction can be seen over the panels, with the cluster apparent size reducing by half over a timescale of only $<0.5$ Myr.}
\label{fig:dist}
\end{figure*}

As the SC spirals inward due to dynamical friction, stars are stripped due to the combined tidal forces exerted by the Galactic field and the SMBH. This process, together with the SC compactness, determines the amount of mass brought into the Galactic center.
In our simulation, the SC enters the inner 10 pc after $60$ Myr \citep[e.g. Figure 24 in][]{ASG17}. Beyond this point, our aim is to understand whether, and for how long, the SC debris remains clearly distinguishable from NC members. Once inside the NC's effective radius of $\sim 4$ pc, the dissolution of the cluster proceeds quickly, and within 1 Myr the cluster cannot be considered a self-gravitating system any longer, as can be seen in the surface density maps shown in Figure \ref{fig:dist}.

The stars deposited during the cluster inspiral might maintain signatures of their origin, showing for instance different kinematics compared to the underlying  Galactic center population. The rotation of the NC \citep{feldmeier14}, however, which is not included in our simulation, might erase some of these features. We therefore add a line-of-sight (LOS) velocity component to the data before recovering features at the present time. We consider the following cases: a) no rotation, b) added rotation in the NC only, c) added rotation in both the NC and the SC stars assuming a relative prograde motion, d) added retrograde rotation in both the NC and the SC stars assuming a relative retrograde motion. In the case b the former SC members do not mix kinematically with the NC while cases c and d represent the extreme scenarios in which the SC members experience the full drag of the NC rotation. To reproduce the observational constraints, we assume that the SC is inclined by 108 degrees with respect to the NC's rotation axis.

\begin{figure*}
\centering
\includegraphics[width=0.43\textwidth]{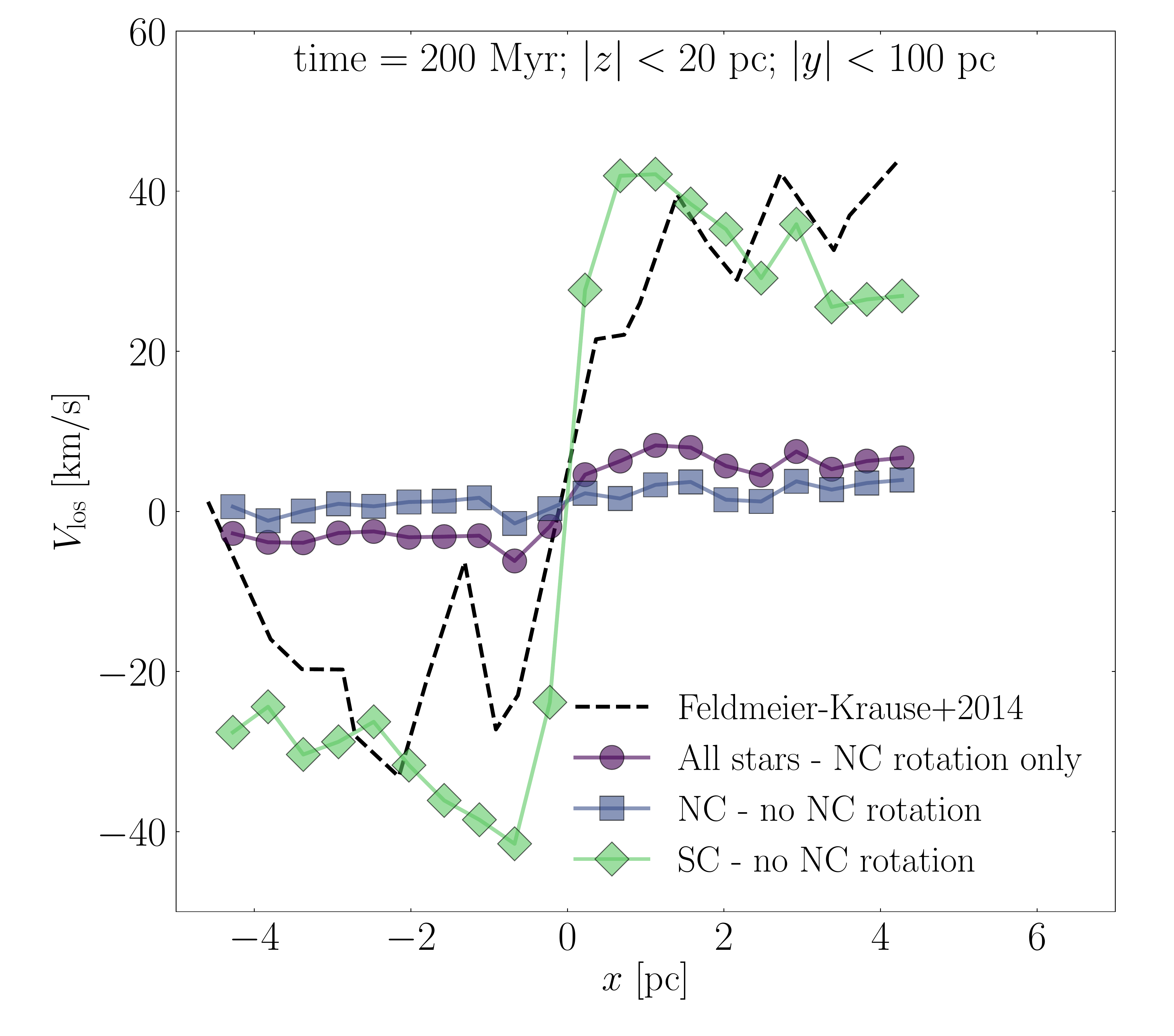}
\includegraphics[width=0.43\textwidth]{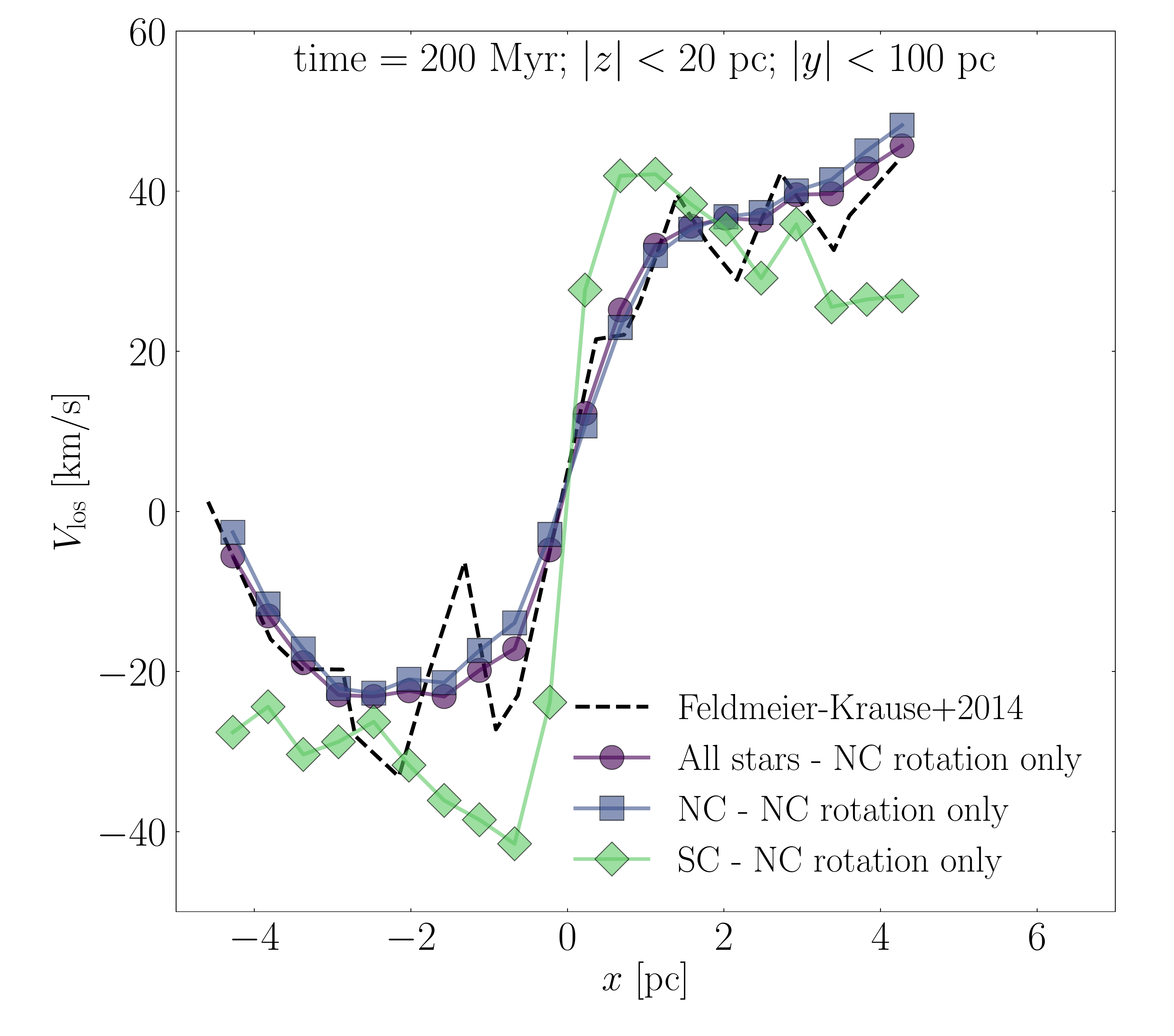}\\
\includegraphics[width=15cm]{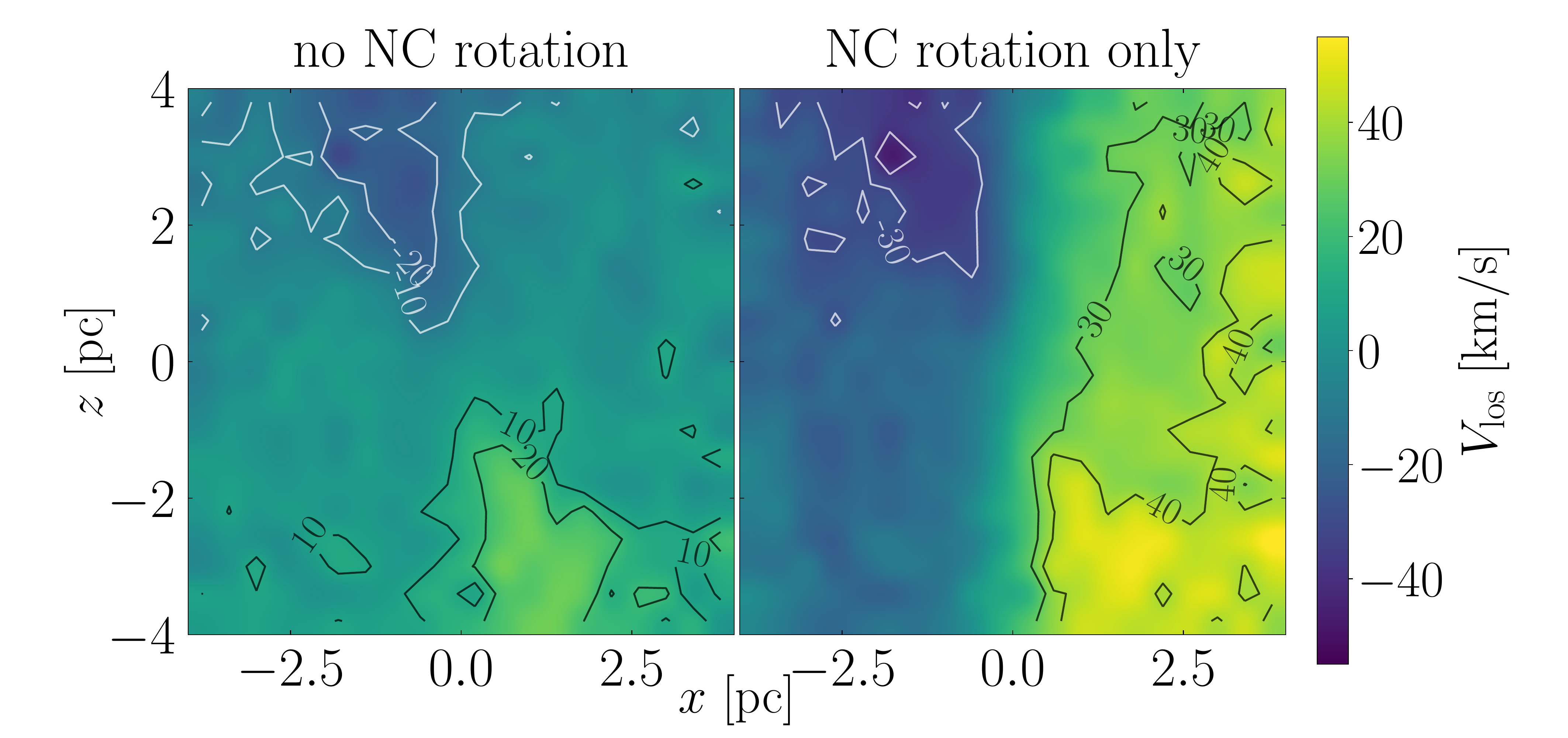}
\caption{Top: line-of-sight velocity profile for stars in our model (points) at time $t=200$ Myr compared to observed  NC stars (dashed line) \citep{feldmeier14}. We show the case of no rotation (left panel) and NC rotation only (right panel). The different lines refer to all stars (circles), NC stars (squares) and SC stars (diamonds) in the region defined by $|x|<10$ pc, $|y|<100$ pc, and $|z|<20$ pc. Bottom: density map of the line-of-sight velocity (assumed to be the $y$ component) for all stars assuming no rotation (left panel) or NC rotation only (right panel). The kinematical signature of the SC is evident as an overdensity tilted by 108 degrees with respect to the NC rotation axis. The density maps are smoothed via a Gaussian kernel. Black and white contours identify the loci of regions characterised by the same LOS velocity. For the sake of visibility, the contours are limited to $[-20,-10,10,20]$ km s$^{-1}$ (left) and $[-40,-30,30,40]$ km s$^{-1}$ (right).}
\label{fig:los}
\end{figure*}

Figure~\ref{fig:los} shows the LOS velocity curve for SC members and background stars separately, as well as for the total combined population, assuming that our LOS lies in the SC's orbital plane, and assuming a field of view of 20 pc in the XZ-plane and of 100 pc along the Y-direction. We compare the non-rotating model (a) and the model in which NC rotates but former SC members orbits remain unaffected by rotation (b).
It is evident that the cluster debris contributes to the overall NC rotation at least within the simulated time, i.e. $\sim 140$ Myr after the SC disruption\footnote{Note that the SC dissolution happens over 65 Myr, whereas the total simulated time is 200 Myr.}, providing a significant contribution to the overall dynamics. The lower panels show the LOS velocity map in both cases. The SC debris appears as a tilted overdensity in the map, even in the case in which a NC rotation component is taken into account. We find that the SC kinematics remains visible even if the SC orbit is retrograde with respect to the NC rotation (see Appendix \ref{app:A}).
We show in Appendix \ref{app:B} that relaxation processes can erase these features over a timescale of at most 3 Gyr.

\begin{figure*}
\centering
\includegraphics[width=\columnwidth]{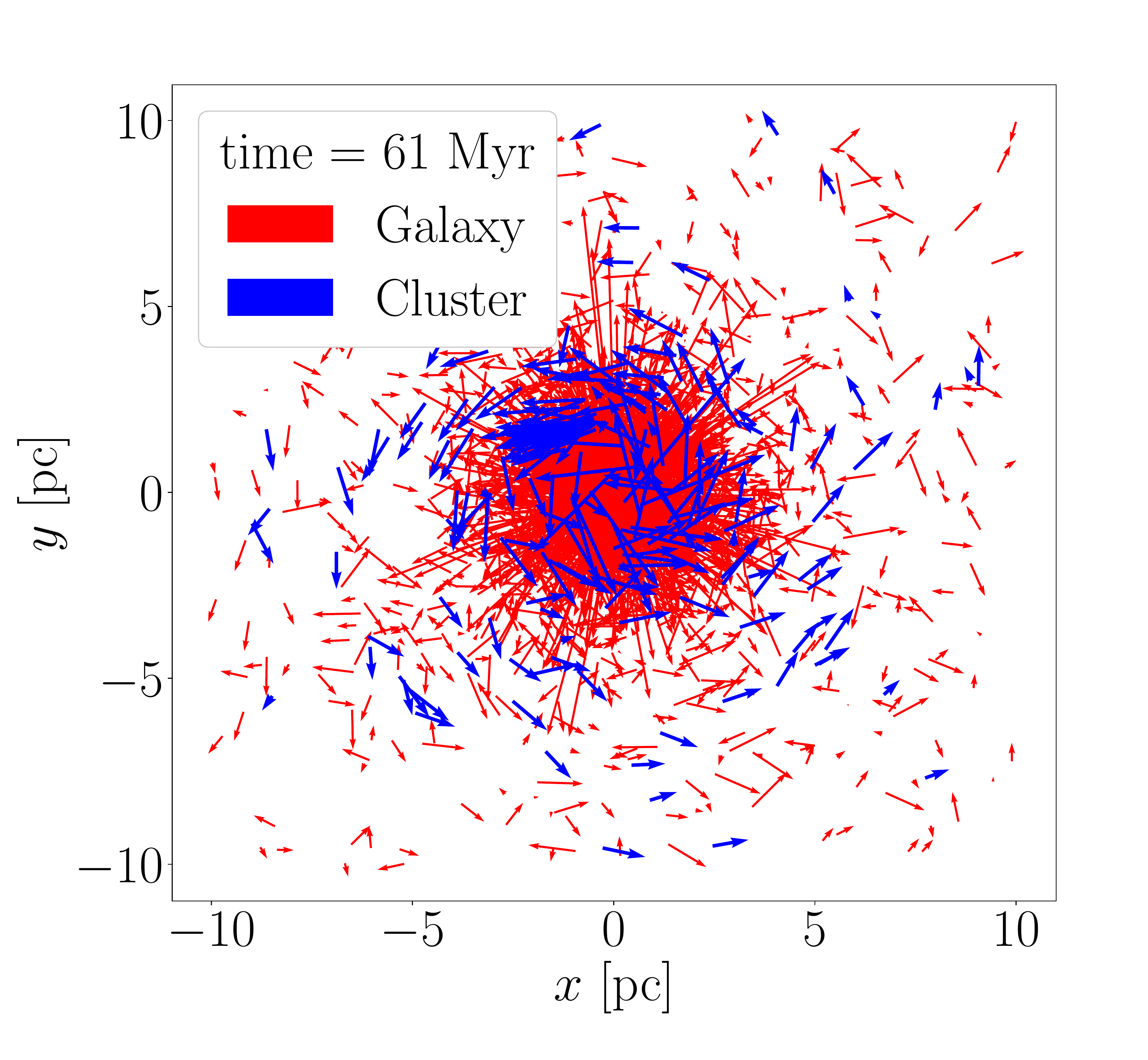}
\includegraphics[width=\columnwidth]{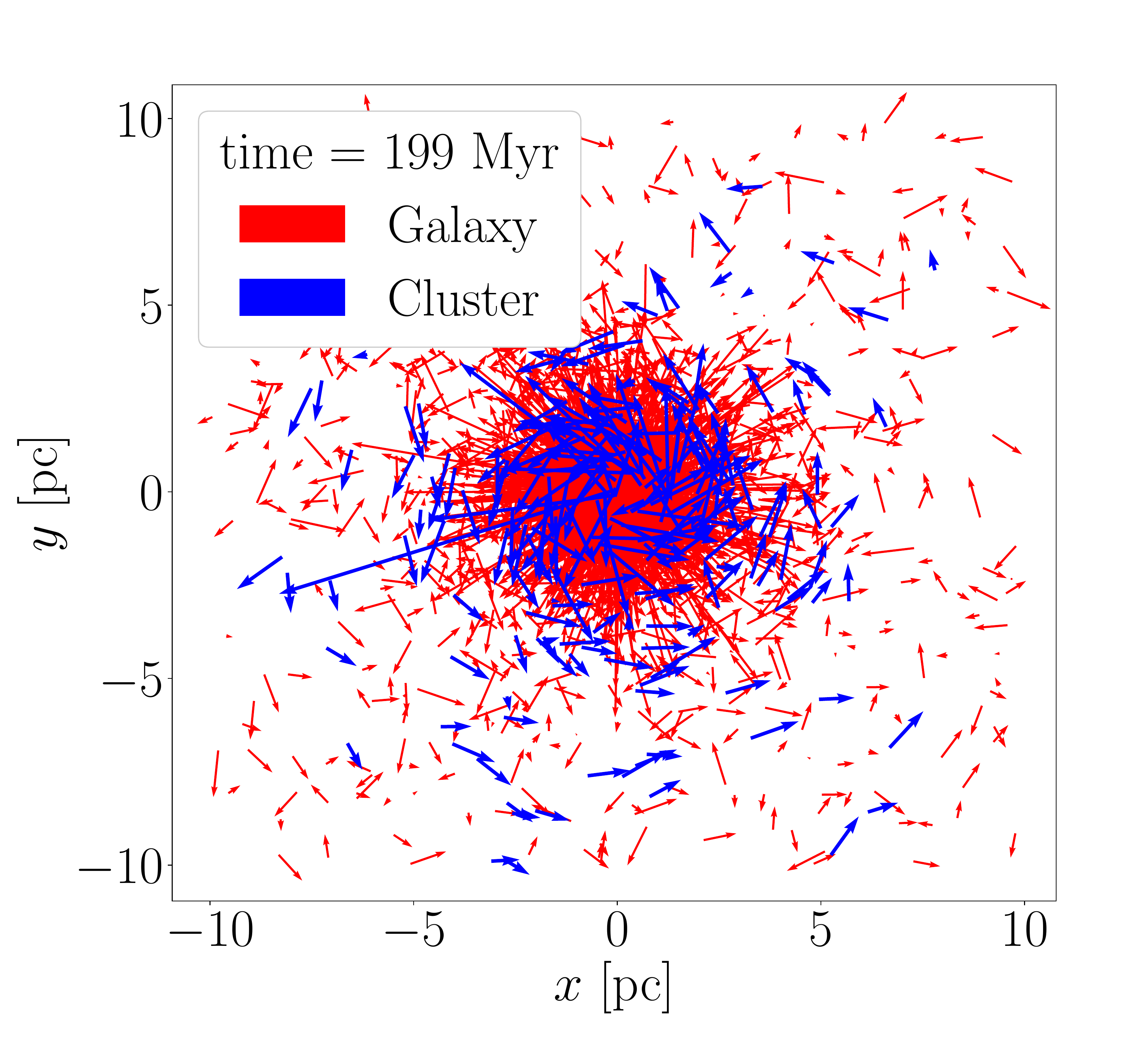}\\
\includegraphics[width=\columnwidth]{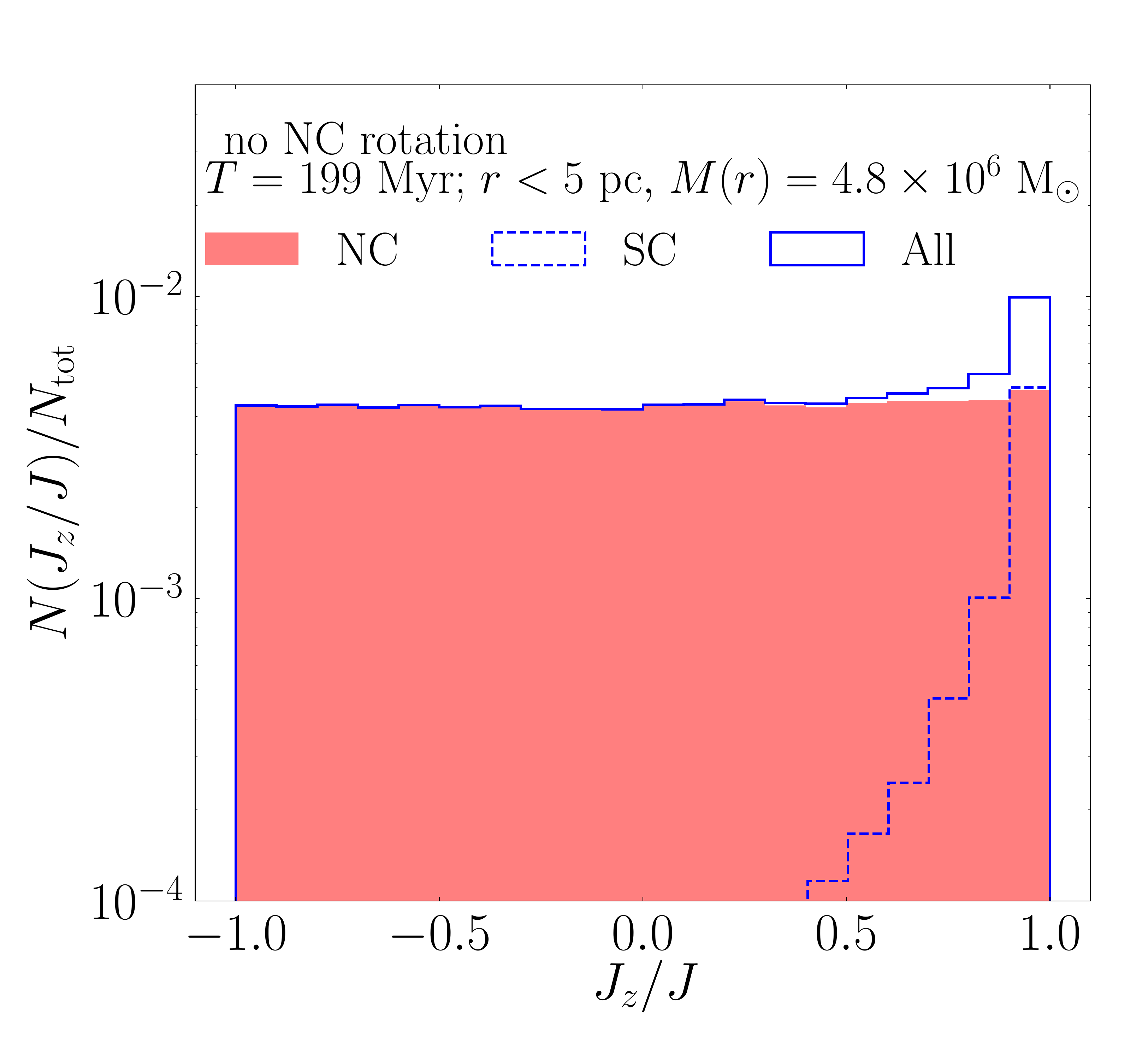}
\includegraphics[width=\columnwidth]{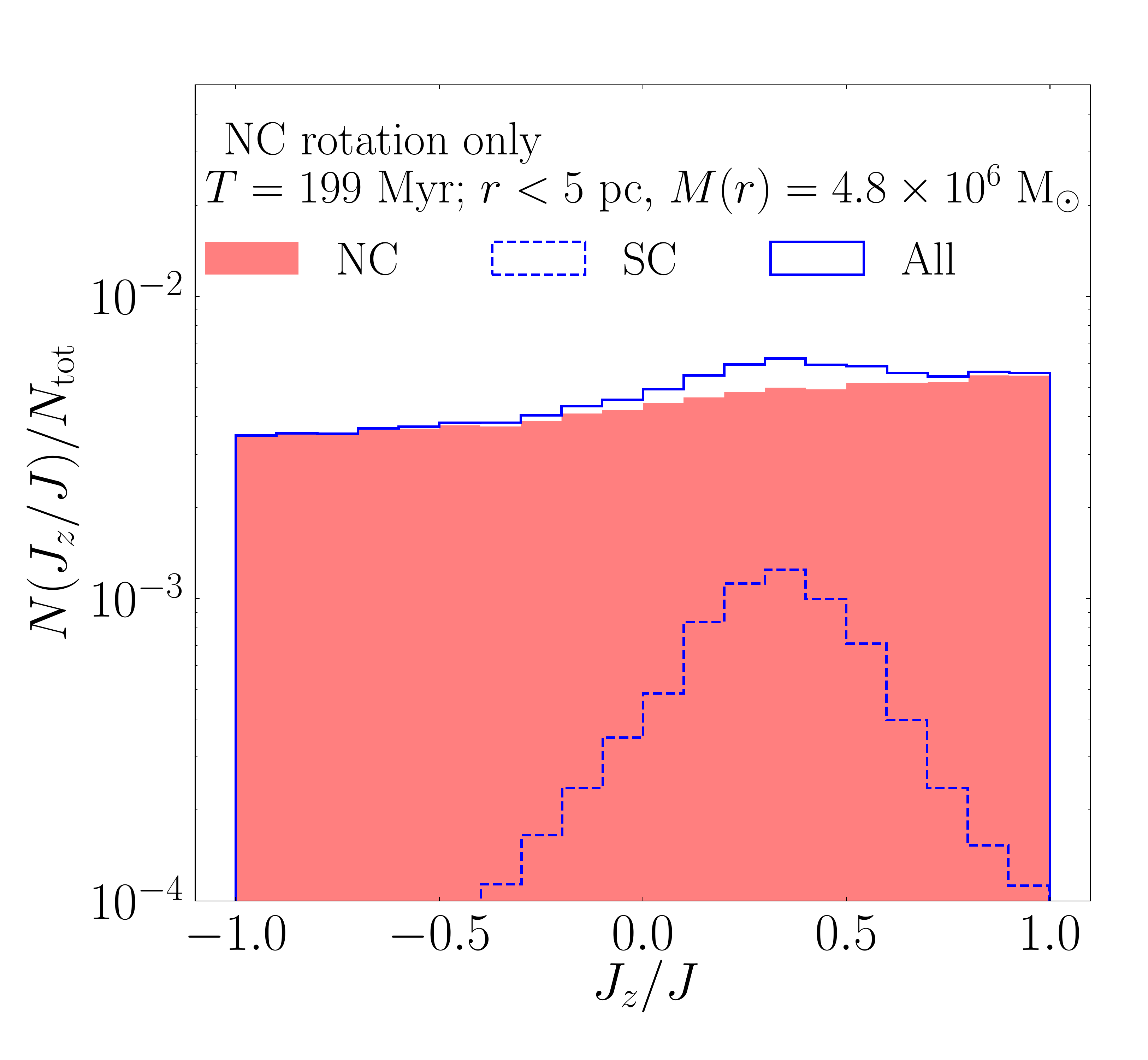}\\
\caption{Top panels: velocity vectors of $2\%$ of the simulated stars ($N_s = 20971$) within a sphere of 5 pc radius centred on the Galactic center assuming no rotation in the NC. Former SC members are labelled with blue arrows, whereas NC stars are represented with red arrows. Vectors' length is proportional to stellar velocity. Bottom panels: distribution of the circularity parameter $J_z/J$  for all stars in the Galactic center (red/filled), the SC (blue/dashed), and the whole system (blue/solid) within a sphere of 5 pc radius centered on the Galactic center assuming a non-rotating (left panel, model a) or rotating (right panel, model b) NC.}
\label{fig:prop}
\end{figure*}

Measuring stellar proper motions for the observed sample would provide further evidence of a recent SC infall. Figure~\ref{fig:prop} shows the velocity vectors of a sample of 20971 stars ($2\%$ of the simulated system) selected assuming $|x|<10$ pc, $|y|<10$ pc, and $|z|<100$ pc and no NC rotation. We find that former SC members constitute $\sim 10\%$ of the sample. Before dissolution, the SC is clearly visible in the map in the form of a concentration in the top-left region of the NC, and remains recognizable for at least 100 Myr after dissolution. The timescale over which these kinematical features persist depends on the process that regulates the relaxation of the SC debris. As discussed in the Appendix \ref{app:B}, two-body relaxation between SC and NC stars erases such features over a timescale of $\lesssim 3$ Gyr, suggesting that the SC infall into the Galactic center is unlikely to have happened earlier than 3 Gyr ago. The effects of the SC debris on the overall kinematics can also be seen in the stellar angular momentum. In particular, for both former SC members and NC stars orbiting inside the inner 5 pc, we calculate the {\it circularity} parameter, defined as the ratio of the z-component of the angular momentum and the total angular momentum of a star, $J_z/J$, and assuming a reference frame coinciding with the directions set by the Galactic center inertia tensor. Values $J_z/J \simeq 1$ indicate a planar prograde motion, whereas $J_z/J \simeq -1$ indicates retrograde motion, and intermediate values indicate a kinematically supported distribution of energies. The bottom panels of Figure~\ref{fig:prop} show the distribution of the circularity parameter calculated in a sphere of radius 5 pc for Galactic and SC stars after time $200$ Myr for cases a) and b).
In the former case we assume that the orbital plane of the SC lies in the X-Y plane, whereas in the latter we assume that the rotation axis of the NC is parallel to the Z-axis. In both cases, SC members determine the formation of a clear peak in $J_z/J$, even when the SC orbit is misaligned with respect to the NC rotation axis.
However, constraining $J_z/J$ requires a level of precision in measuring stellar distances not achievable with current instruments. An alternative, measurable quantity would be the ``sky-projected angular momentum'' \citep[see Equation B1 in][]{paumard06}, which combines stellar position and velocities lying in the plane normal to the LOS. However, we find that if our LOS -- which defines the z-direction -- lies in the NC rotation plane it is practically impossible distinguishing SC debris from the overall stellar population. 

To enable a comparison with future observational data, we provide position, velocity, and angular momentum of all stars in our simulation as detailed in Section \ref{data}.

\subsection{Evolution of the Metallicity Distribution via Star Cluster Infall}

\begin{figure}
\centering
\includegraphics[width=\columnwidth]{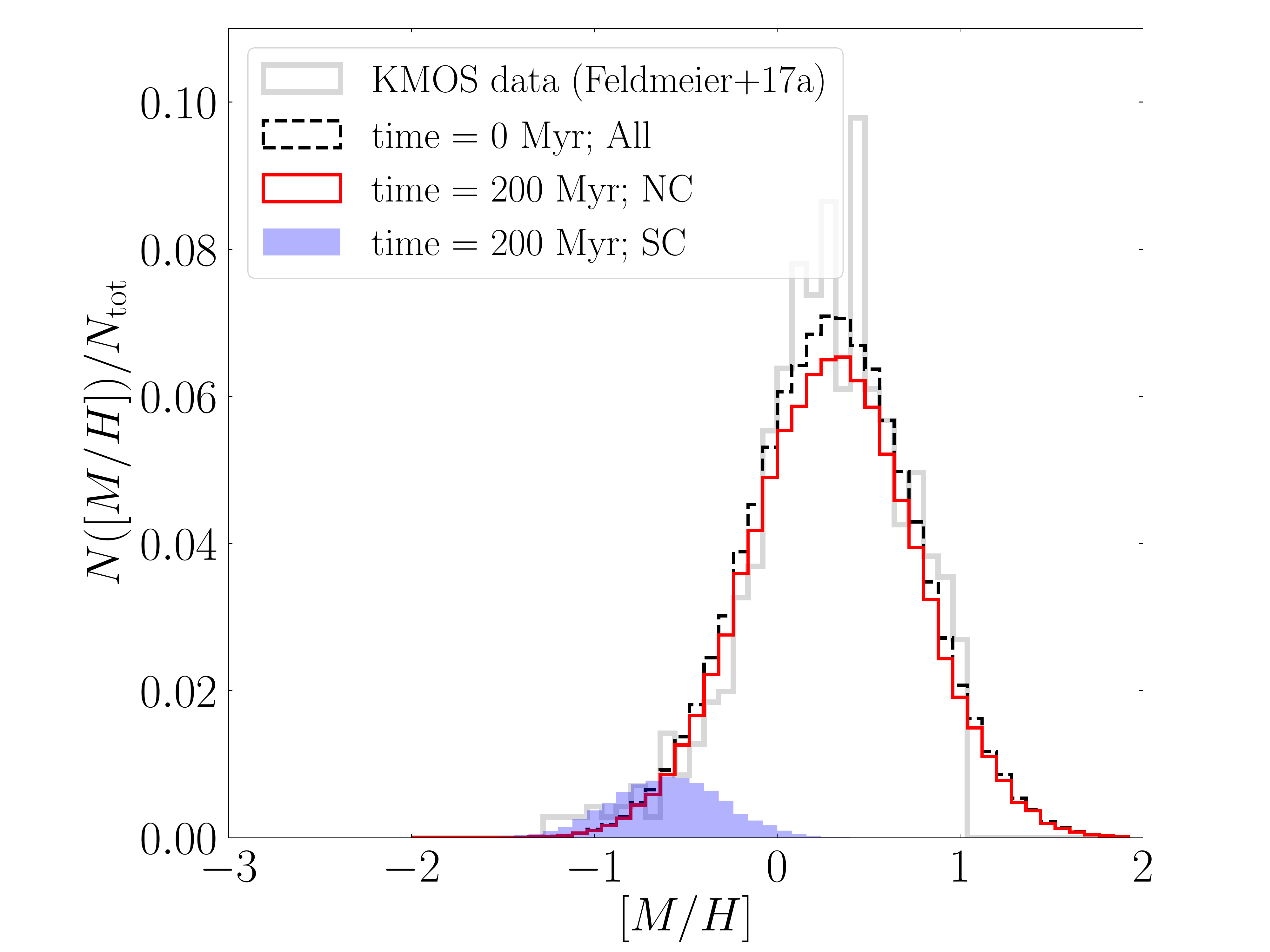}\\
\caption{Metallicity distribution for the NC members at time $t=0$ (black dotted line) and for both NC (red empty histogram) and SC (blue filled histogram) members at time $t=200$ Myr, in the innermost 5.5 pc. The observational data from KMOS (see \citet{feldmeier17a}) exploited in our companion paper are overlayed to the model (light grey histogram).}
\label{fig:met}
\end{figure}

A further peculiarity of the stars discussed in our companion paper (Do et al., in press) is that they are considerably more metal poor than the overall distribution. To test whether this finding is compatible with a SC infall event, we adopt the following ``tagging'' procedure: i) at the start of the simulation, at time $t=0$, we tag each star in the galaxy and in the SC with a $[M/H]$ value assigned according to a given distribution, ii) at time $t$, we reconstruct the metallicity distribution of stars in a given region of phase space. We follow the observational constraints of a metal-rich population with a broad $[M/H]$ distribution peaked at $[M/H]\sim 0.33$ and  a metal-poor population with a weak peak at $[M/H] \simeq -0.54$ and model the metallicity distribution with two Gaussian curves. For SC members, we assume that the distribution has a mean value of $[M/H] = -0.6$ and a spread of $\sigma = 0.5$, whereas for NC stars we set $[M/H] = 0.3$ and $\sigma = 1.5$. Figure~\ref{fig:met} shows the reconstructed $[M/H]$ distribution for stars in the central 5.5 pc at times $60-200$ Myr. Under our simplistic assumptions, the overall distribution of $[M/H]$ resembles the observed one (see Figure 3 in Do et al.), with former SC stars dominating the distribution in the range $[M/H] = (-1,0)$.  We note that the relative amount of metal-poor versus metal-rich stars is completely determined by the competing actions of dynamical friction, which brings SC stars toward the Galactic center, and tidal disruption, which tends to strip stars away from the SC. Inside the innermost 4 pc, we find that former SC members constitute $\sim 7.36\%$ of the total stellar population, compatible with the percentage found in the observed sample ($\sim 7\%$). The accumulation of such a large number of SC members into the innermost NC regions is due to the fact that the cluster core survives the intense tidal forces and is therefore able to deposit stars into the NC.

\section{Discussion}
\label{sec:dis}

\subsection{The Origin of the Metal-Poor Stellar Population: Galactic or Extragalactic?}
\label{sec:origin}

To place constraints on the origin of the SC progenitor we develop a semi-analytic approach, described in Appendix \ref{app:C}, to model the evolution of the SC before it reaches the Galactic center under the effects of dynamical friction and tidal disruption processes. We focus on the two special cases where the SC is either the remnant of a Galactic cluster (model "GAL") or the nucleus of a dwarf galaxy (model "EXT"). We create two samples of 50,000 SC {\it progenitor} models each, varying the orbital and structural properties of the infaller as detailed in Appendix \ref{app:C}.

\begin{figure*}[h]
\includegraphics[width=0.33\textwidth]{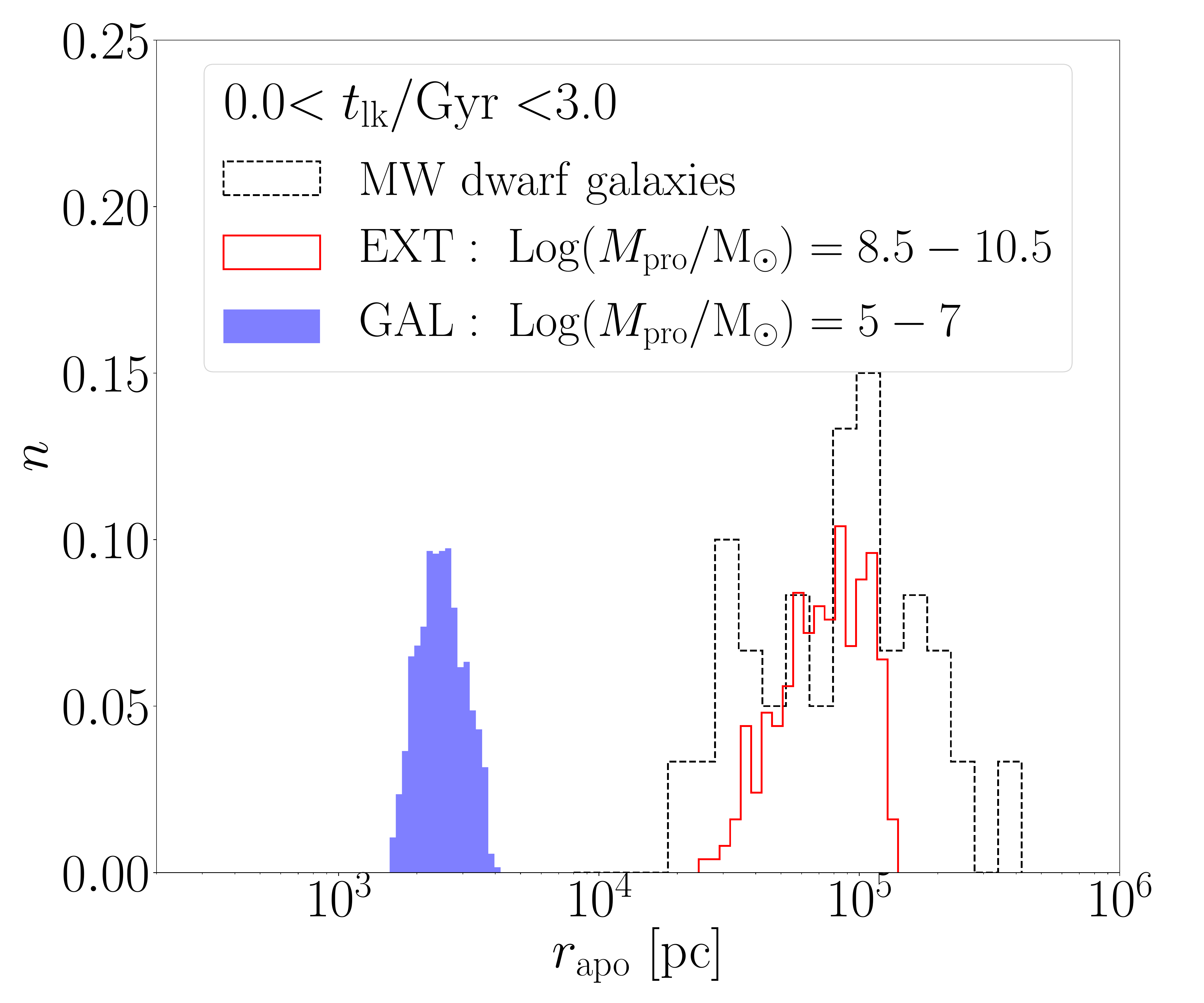}
\includegraphics[width=0.33\textwidth]{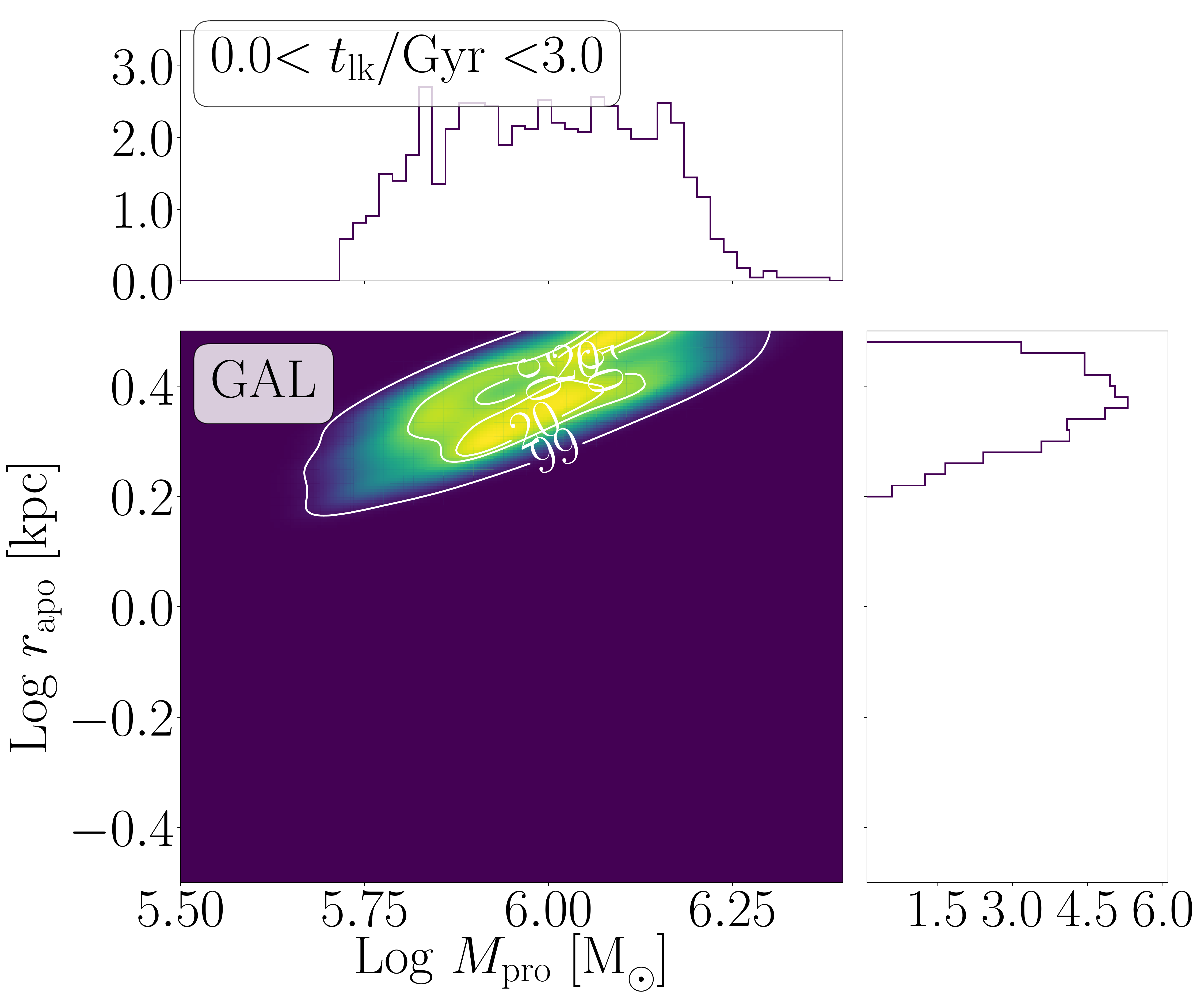}
\includegraphics[width=0.33\textwidth]{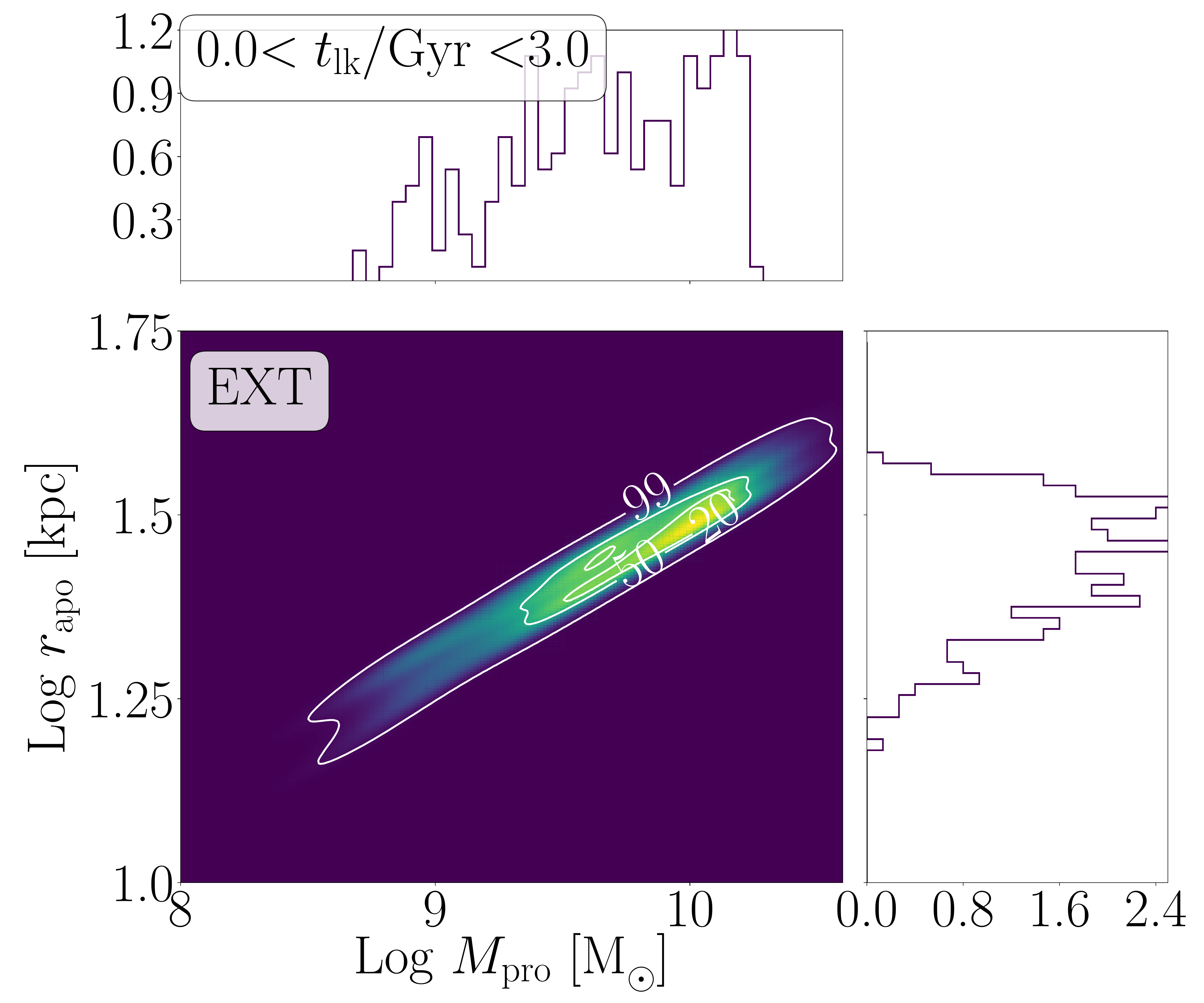}\\
\includegraphics[width=0.33\textwidth]{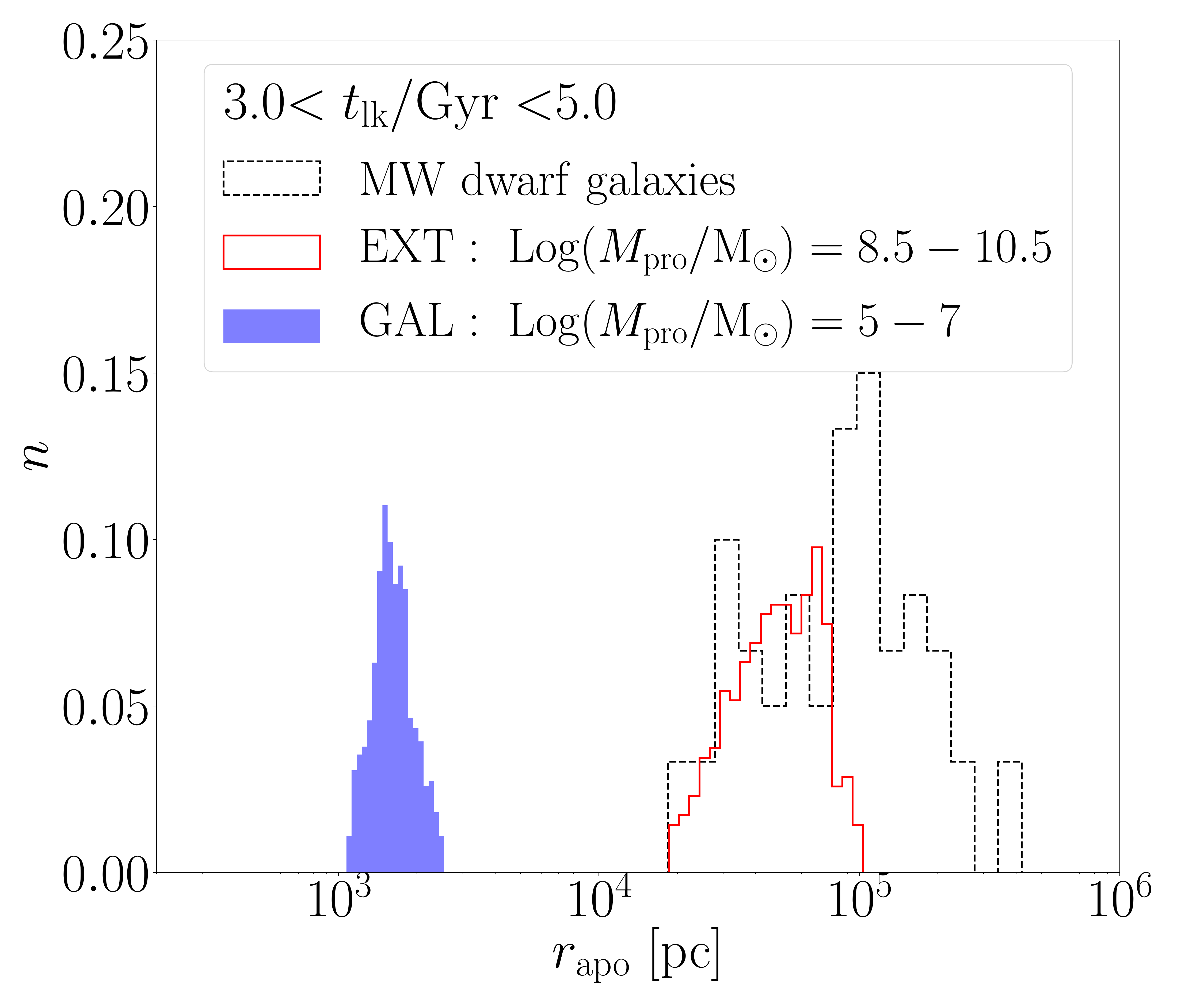}
\includegraphics[width=0.33\textwidth]{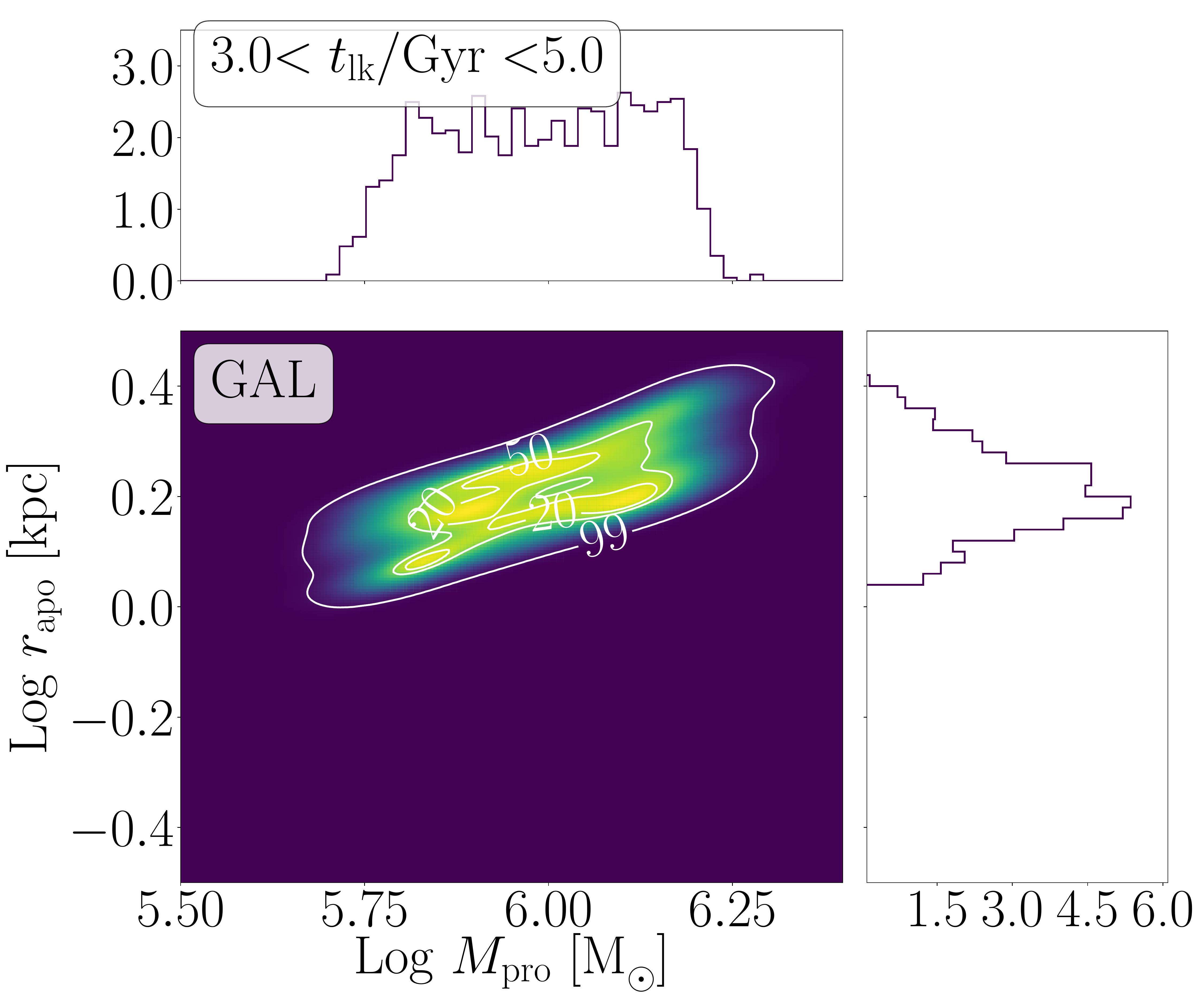}
\includegraphics[width=0.33\textwidth]{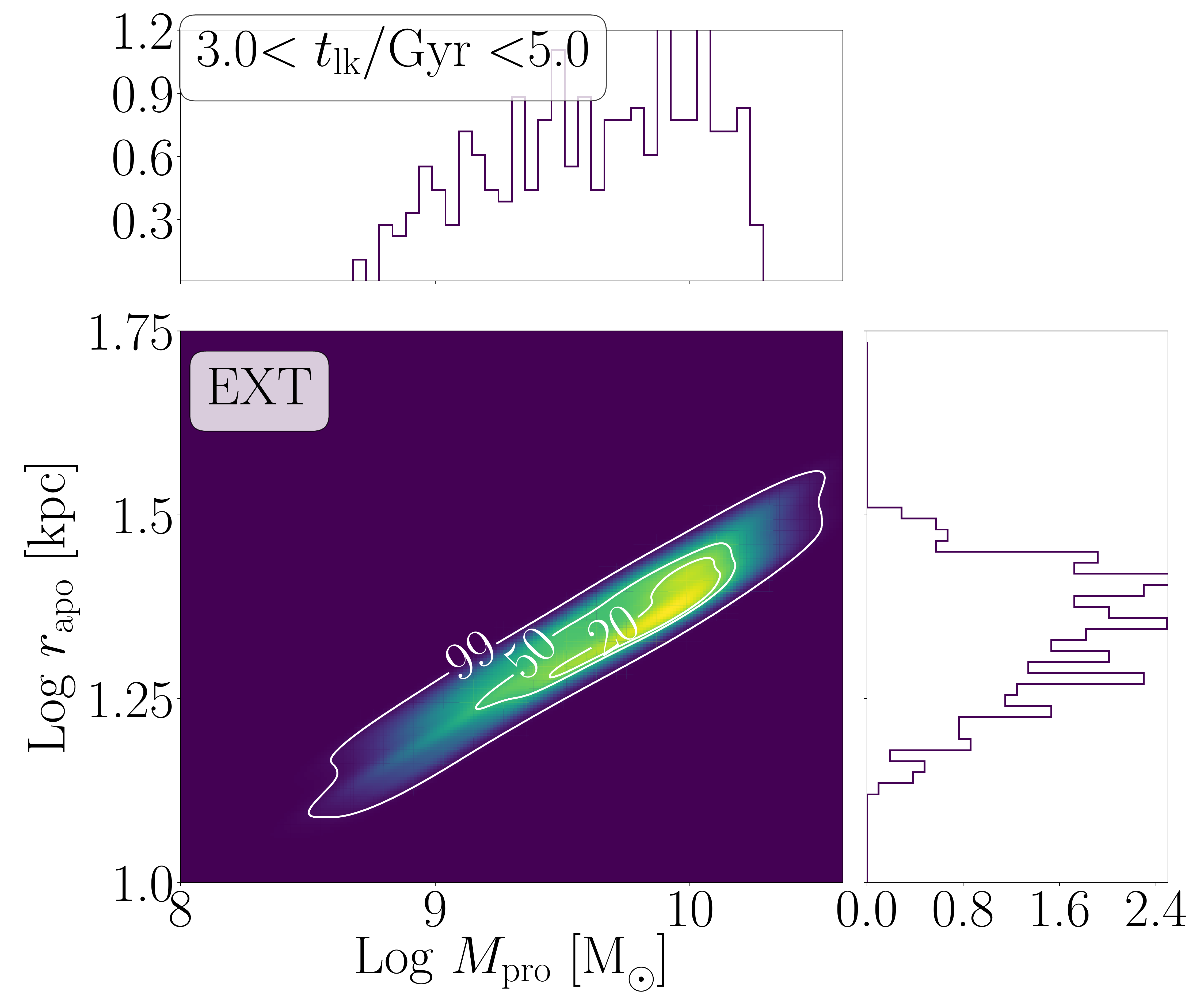}\\ 
\includegraphics[width=0.33\textwidth]{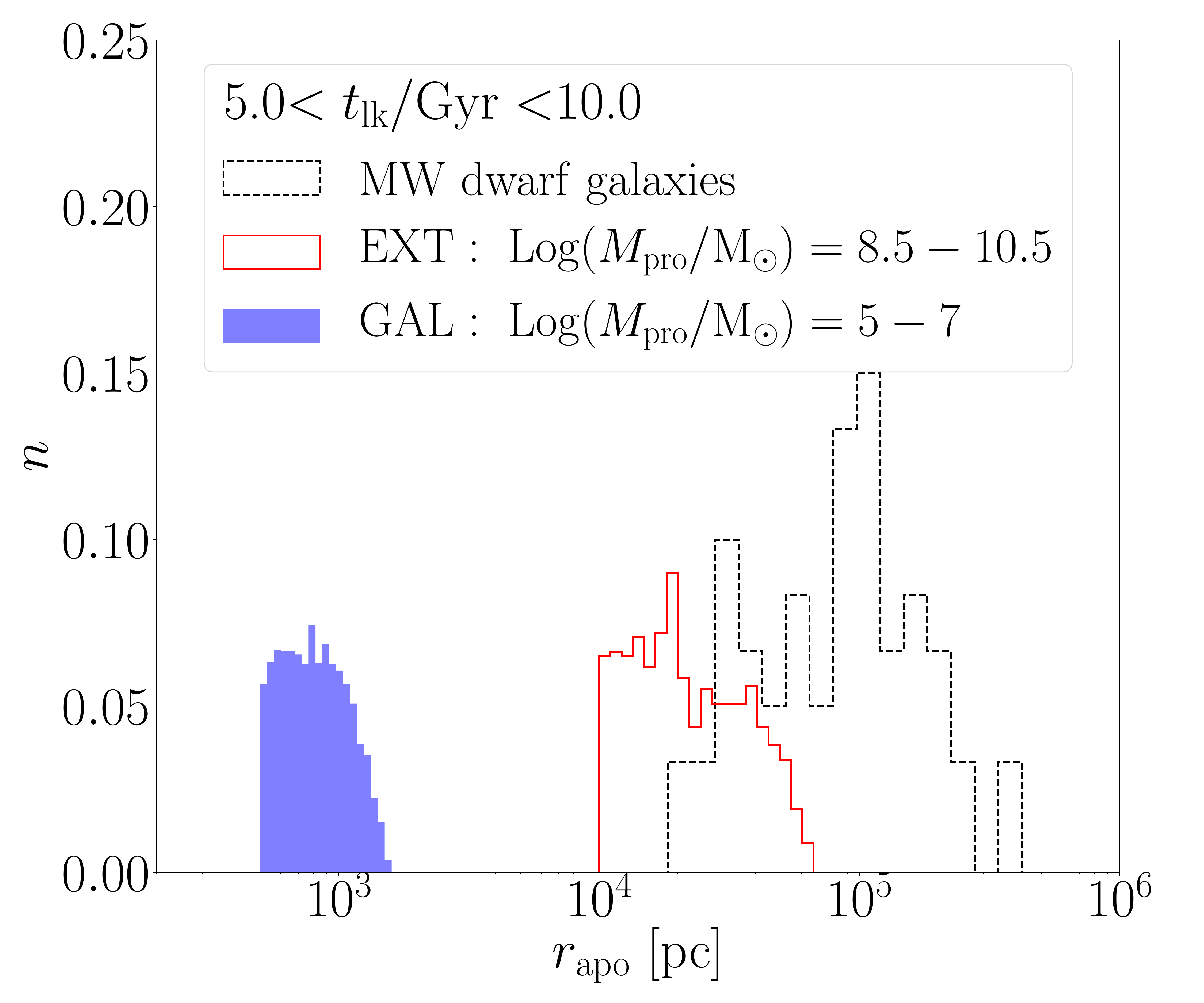}
\includegraphics[width=0.33\textwidth]{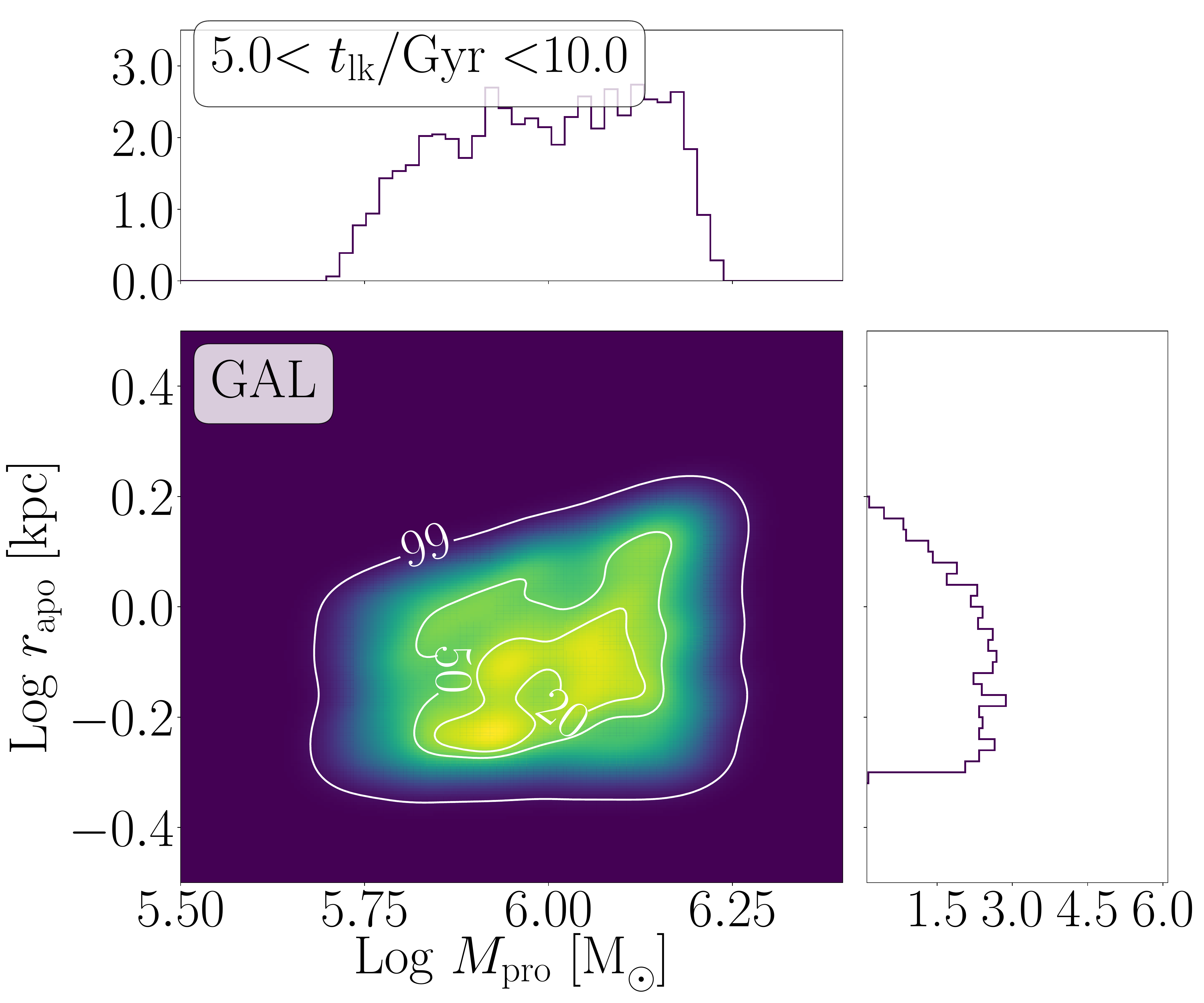} 
\includegraphics[width=0.33\textwidth]{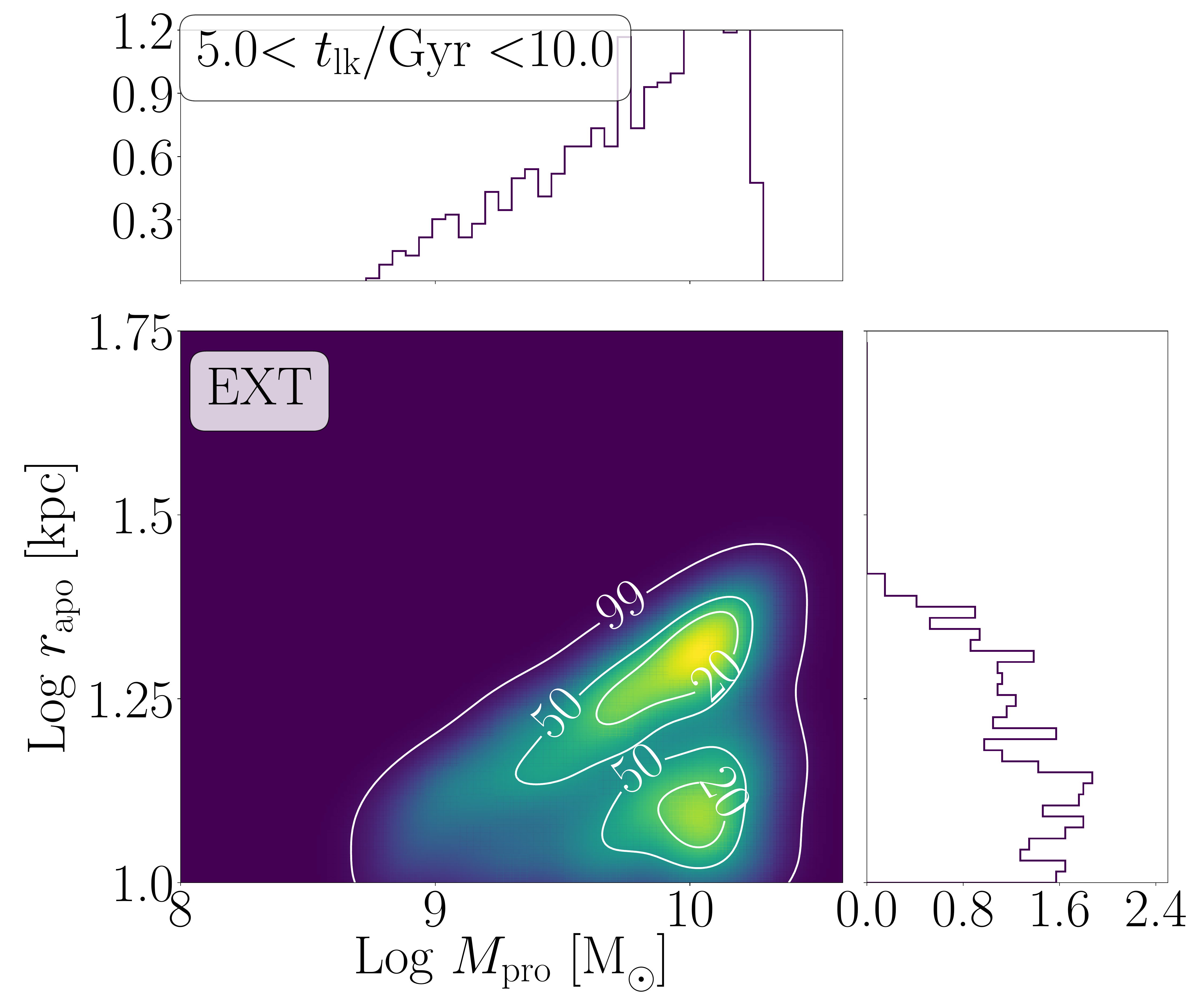} 
\caption{Left panels: location of the SC birth-site assuming a Galactic (GAL, blue filled histogram) or extra-galactic (EXG, red empty histogram) origin. From top to bottom, panels correspond to infall time intervals $(0-3),~(3-5),~(5-10)$ Gyr ago, respectively. We include in the plot the distribution of current locations of MW satellites (black dashed histogram). All histograms are normalized to the total number of objects in each sample. Central(right) panels: combined surface distribution of initial mass and galactocentric distance in the case of a galactic(extragalactic) origin, assuming the same infall time ranges as in the left panels.}
\label{fig:rori}
\end{figure*}

Among all models, we select progenitors delivering $(0.5-1.5)\times 10^6\Ms$ in stars in the central $50$ pc\footnote{Our semi-analytic treatment of the infall process does not account for tidal effects arising from the NC, thus we limit the model to distances larger than 50 pc where the NC's tidal field is negligible compared to the galactic field.}, so to be compatible with the simulated SC. We store their initial position, mass, and {\it arrival time} $t_{\rm lk}$, namely the time since the cluster crossed the 50 pc radius. We differentiate between {\it late} ($t_{\rm lk} < 3$ Gyr), {\it recent} ($3 < t_{\rm lk}/{\rm Gyr} < 5$), and {\it early} infalls ($5 < t_{\rm lk}/{\rm Gyr} < 10$). Since we assume that the progenitor system formed 10 Gyr ago, the arrival time represents a measure of the time that the SC debris spent at the Galactic center: an arrival time of 3 Gyr means that the SC formed 10 Gyr ago and it took 7 Gyr to reach the inner 50 pc. We highlight the fact that while our numerical model focuses on a SC with total mass $10^6\Ms$, having such a mass is not a {\it sine qua non} condition. 
A cluster remnant that is lighter but denser (or heavier but sparser) than the one used in our model could bring, in principle, the same amount of mass into the Galactic center. 
The left panels of Figure~\ref{fig:rori} show the distribution of initial Galactocentric distances $r_{\rm apo}$ for both GAL and EXT models under the assumptions above. If the metal-poor population is the remainder of a Galactic cluster, it should have formed at $r_{\rm apo} \sim 3-5$ kpc, whereas in the case of a spiralling dwarf galaxy we find $r_{\rm apo} = 10-100$ kpc. For comparison, we also show the current location of the MW dwarf satellite galaxies\footnote{\url{https://web.archive.org/web/20140219170336/http://www.astro.uu.se/~ns/mwsat.html}} in the figure. We note that the range of initial locations for late and recent infalls ($t_{\rm lk}<3-5$ Gyr) is shifted toward larger values compared to early infall events, since the farther the progenitor the longer the dynamical friction timescale\footnote{For simplicity we assume that SC progenitors formed 10 Gyr ago, as typical for Galactic globular clusters.}.  
The right panels of Figure \ref{fig:rori} show the combined distribution of $r_{\rm apo}$ and $M_\gc$ for model GAL and EXT and for different $t_{\rm lk}$ values. 

\begin{figure}
    \centering
    \includegraphics[width=0.97\columnwidth]{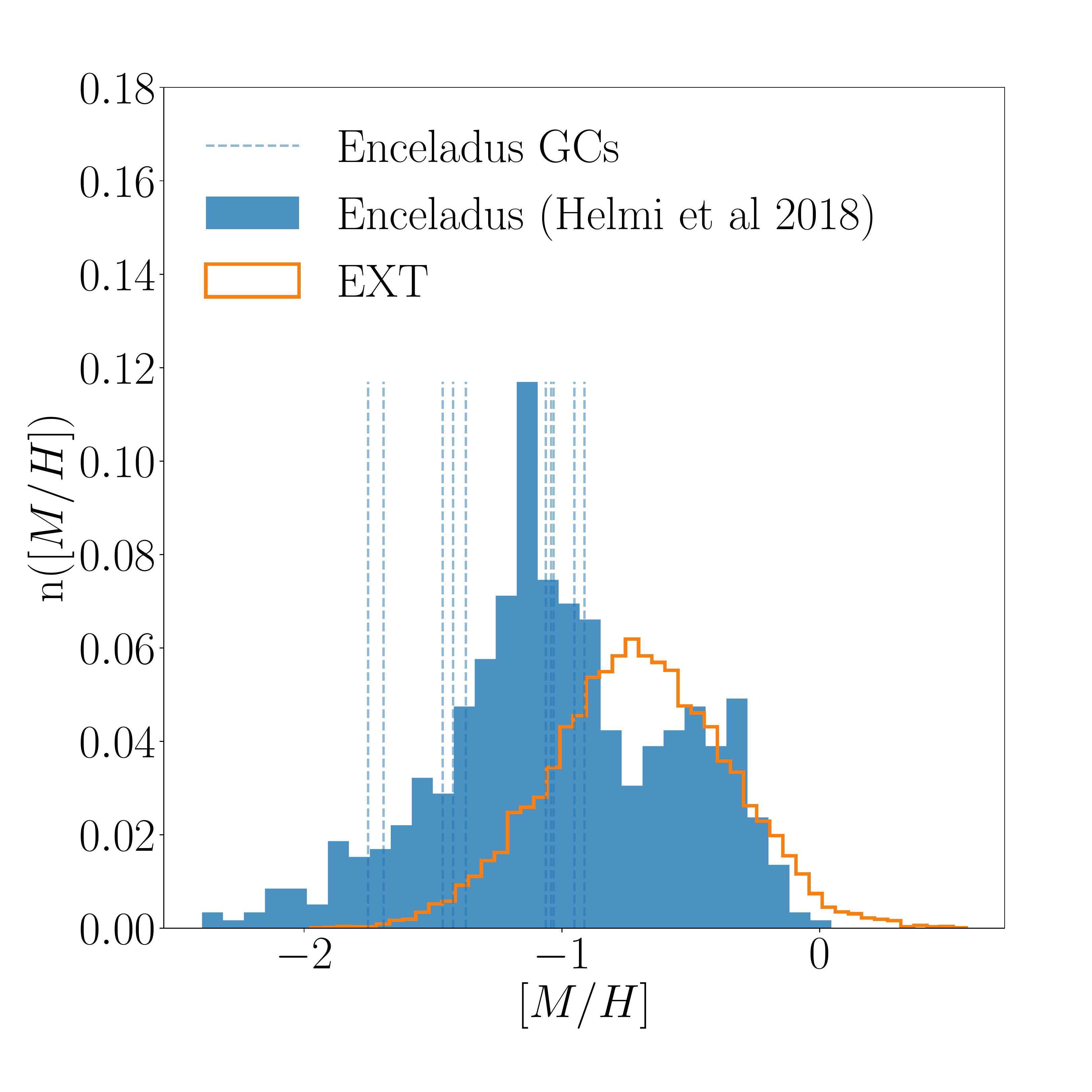}\\
    \includegraphics[width=0.97\columnwidth]{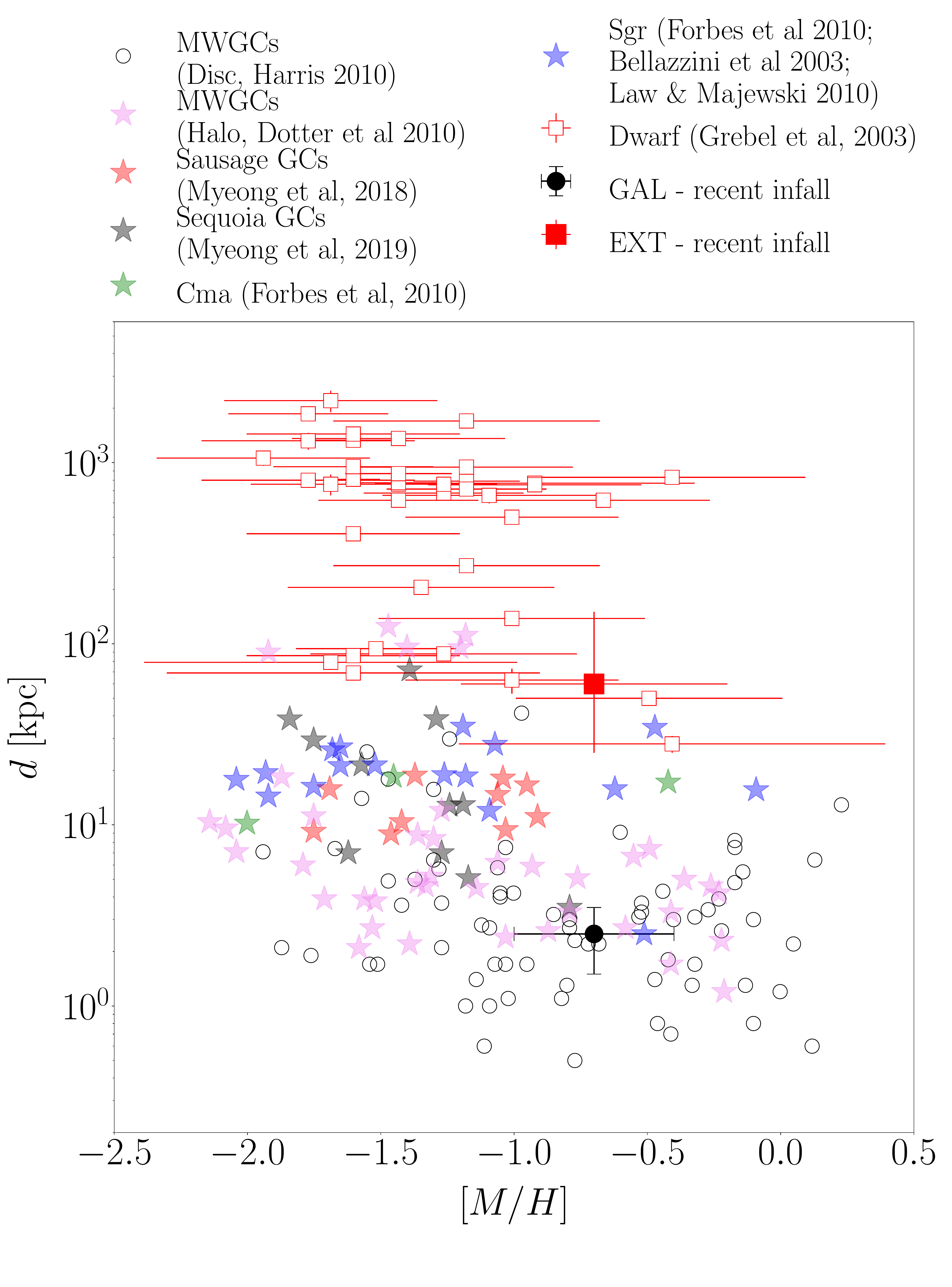}
    \caption{Top panel: metallicity distribution for the EXT model (orange empty histogram) and for the Enceladus stream (blue filled histogram). Bottom panel: Galactocentric distance versus metallicity $[M/H]$ for GAL (black filled dot) and EXT (red filled square) models, globular clusters in the MW disc \citep[black empty dots][]{harris10} and in the halo \citep[pink stars,][]{dotter10}, potentially accreted clusters that might belong to Enceladus \citep[red stars,][]{myeong18}, Sequoia \citep[black stars,][]{myeong19}, the Sgr dwarf \citep[blue stars,][]{bellazzini03,law10}, or the Canis Major (Cma) structure \citep[green stars,][]{forbes10}, and MW dwarfs \citep[red empty squares,][]{grebel03}.} 
    \label{fig:comp}
\end{figure}

Another intriguing possibility is an hybrid EXT/GAL scenario in which a dwarf galaxy accreted into the MW halo left one of its star clusters sufficiently close to the Galactic center ($ \lesssim 1-3$ kpc) for it to spiral-in within a Hubble time.
In fact, a number of structures in the Milky Way have been recently identified as possible relics of accretion events, e.g. the Sausage \citep[Gaia Enceladus,][]{belokurov18,myeong18}, likely the remnant of a dwarf galaxy that interacted with the MW $\sim 10$ Gyr ago \citep{helmi18,DiMatteo19}, or Sequoia \citep{myeong19}. These are likely remnants of satellites as massive as $10^{10} - 10^{11}\Ms$ that polluted the Milky Way's star cluster population with their own clusters. Thus, in the case of an EXT or GAL+EXT origin, the metallicity of the progenitor should be comparable to that of dwarf galaxies orbiting the Milky Way or of accreted clusters. We compare the metallicity distribution of our EXT model with that of Gaia-Enceladus stars \citep{helmi18} in the top panel of Figure ~\ref{fig:comp}, which shows that the Galactic center stellar population is more metal rich than Gaia-Enceladus stars. 
This rules out, or at least disfavors, the scenario in which the nucleus of Enceladus reached the Galactic Centre, unless it was significantly more metal-rich than the overall metallicity of the system. Note that the average metallicities of Enceladus/Sausage globular clusters are lower than the peak metallicity measured for the structure at the Galactic Centre.

The bottom panel shows a comparison of  the Galactocentric distance versus metallicity\footnote{ To convert the measured $[Fe/H]$ into $[M/H]$, we fit the $[\alpha/Fe]$-$[Fe/H]$ data provided by \cite{helmi18} using a linear relation.} our EXT and GAL progenitors with the sample of dwarf galaxies compiled by \cite{grebel03}, as well as Galactic globular clusters and Sequoia and Sausage clusters. Our EXT model is located in an area of the relation scarcely populated by MW satellites, although this dearth of members might be due to the fact that dwarf galaxies were accreted a long time ago. 
In the case of a Galactic origin and a recent infall, the progenitor location is $r_{\rm apo} \simeq 3-5$ kpc with $[M/H]\sim -0.7$. In the same panel, we compare our GAL model with Galactic globular clusters\footnote{To convert the $[Fe/H]$ provided by \cite{harris10} we assume a $[\alpha/Fe]=0.3$ \citep{pritzl05}} \citep{harris96,harris10}. The plot shows a remarkably good agreement between our model and low-orbit globular clusters. All the closest clusters in the distance-metallicity plane shown here (e.g. NGC6569, NGC6256, NGC6171, NGC6401, and NGC6638) have masses $\sim 1-3\times 10^5\Ms$, thus smaller than the expected progenitor mass, but these clusters might have undergone mass loss over their 10 Gyr lifetime.

To test the feasibility of the GAL+EXT scenario, we highlight halo GCs \citep{dotter10}, GCs associated with the Sequoia and Enceladus streams \citep[see][and reference therein]{myeong18,myeong19}, the Sagittarius  \citep[Sgr,][]{bellazzini03,forbes10,dotter10,law10} and Canis Major dwarf galaxies \citep[Cma][]{forbes10}. We find that only two GCs associated with Sequoia and Sgr are compatible with our model, while the range of distances and metallicities turn out to be compatible with both disc and halo GCs. Since the latter are expected to be accreted during the MW build-up (at least a sub-sample of them) we cannot rule out an extra-galactic origin for the metal-poor population at the Galactic center.

Nonetheless, the Galactic scenario seems to provide a better match to the observational constraints, and is therefore our favored explanation for the origin of the peculiar metal-poor stars at the Galactic center. The progenitor cluster could have formed in-situ, as the comparison with low-orbit globular clusters suggests, although an accretion origin is plausible in light of the possible accretion of low-orbit halo GCs during the MW assembly.

\section{Conclusions}
\label{sec:con}
We explore an infall scenario as the possible origin of a population of metal-poor, fast rotating stars observed in the innermost 2 pc of the Galactic center. We use state-of-the-art $N$-body simulations to model the orbital decay of a stellar system with mass $10^6\Ms$ located at an initial distance of $50$ pc from the Galactic center, taking into account the gravitational effects from the infalling cluster, the NC, and the central SMBH. Our results can be summarized as follows:
\begin{itemize}
    \item the inspiral happens over a timescale of $\sim 60$ Myr, after which tidal forces lead to cluster disruption in $\sim 1-10$ Myr; 
    \item once inside the NC, the cluster debris preserves a clear kinematical signature, namely a higher level of rotation compared to the NC population, which is particularly evident in the line-of-sight velocity profile. This supports an infall origin for the observed metal-poor stars;
    \item former cluster stars ($\sim 7.2\%$ of the stars in the inner 4 pc) have velocity vectors that draw a clear pattern in the cluster's orbital plane and
    a distinct distribution of the {\it circularity} parameter;
    \item the reconstructed metallicity distribution obtained by tagging stars in the simulation with a metallicity sampled from the observed distribution reveals a clear contribution of cluster members at low $[M/H]$ and is fully consistent with observations;
    \item we model the effects of dynamical friction and tidal disruption of possible galactic and extra-galaactic cluster progenitors with a semi-analytic model, distinguishing among early, recent and late inspiral.
    For a Galactic progenitor, our models suggest an initial location at $r_{\rm apo} = 2-5$ kpc for late infall and $r_{\rm apo} = 0.5-8$ kpc for early infall, whereas for an extragalactic progenitor we find $r_{\rm apo}$ from $30-120$ kpc (late infall) to $10-100$ kpc (early infall);
    \item a comparison with known accreted structures and their clusters, as well as Galactic globular clusters, shows no connection with Gaia-Enceladus, Sequoia or the Sagittarius dwarf but is compatible with low orbit clusters either in the disk or the halo. 
	\end{itemize}
    
We conclude that the most likely scenario for the observed metal poor population is the remnant of a star cluster formed in the inner regions of the Galaxy, although the possibility that it has been deposited by a disrupting dwarf galaxy 10~Gyr ago cannot be fully ruled out. The identification of footprints from such a disrupted satellite would shed light on the origin of the system.

\section*{Acknowledgements}
The authors are warmly grateful to the referee for their insightful and detailed report which helped to improve the manuscript.
MAS acknowledges financial support from the Alexander von Humboldt Foundation for the research program ``The evolution of black holes from stellar to galactic scales'' and through the Volkswagen Foundation Trilateral Partnership project No. I/97778 ``Dynamical Mechanisms of Accretion in Galactic Nuclei''. MAS and NN gratefully acknowledges support by the Deutsche Forschungsgemeinschaft (DFG, German Research Foundation) -- Project-ID 138713538 -- SFB 881 (``The Milky Way System'', subproject Z2 and B8). Part of the simulations presented in this work have been performed with the bwForCluster of the Baden-W\"{u}rttemberg's High Performance Computing (HPC) facilities thanks to the support provided by the state of Baden-W\"urttemberg through bwHPC and the German Research Foundation (DFG) through grant INST 35/1134-1 FUGG.

\appendix

\section{The effects of nuclear cluster rotation}
\label{app:A}

In order to asses the importance of the NC pristine rotation in washing out the kinematic features of the metal-poor population we post-processed our data following four different sets of assumptions, limiting our analysis to particles in a box of 5 pc length and assuming particles with depth $< 100$ pc:

\begin{itemize}
\item[i)] the NC is non-rotating;
\item[ii)] a los velocity component is added to both NC and SC particles ;
\item[iii)] a los velocity component is added to NC particles only;
\item[iv)] a los velocity component is added to NC particles only, assuming that the rotation is retrograde compared with the SC orbital motion.
\end{itemize}

\begin{figure}
\includegraphics[width=0.45\textwidth]{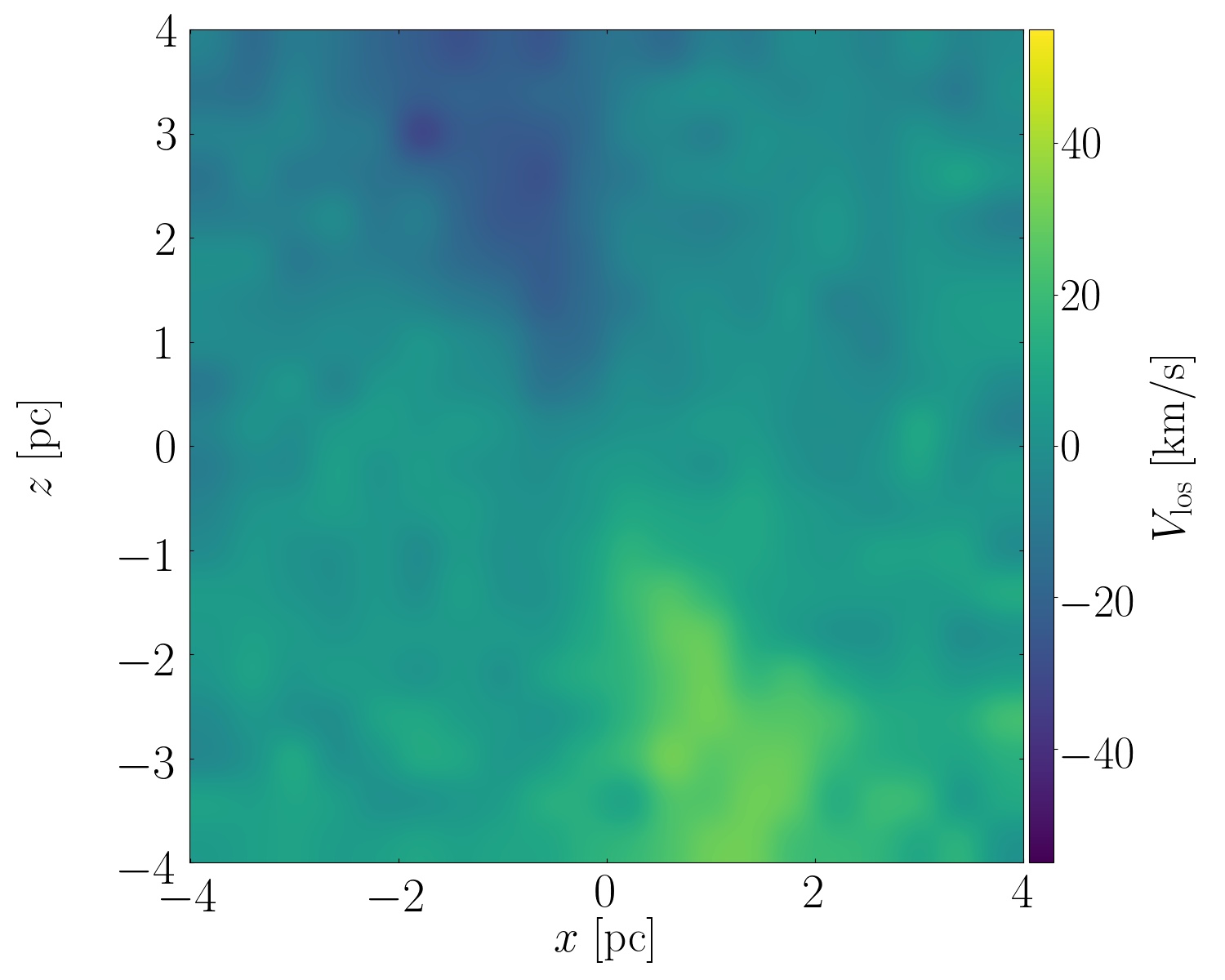}
\includegraphics[width=0.45\textwidth]{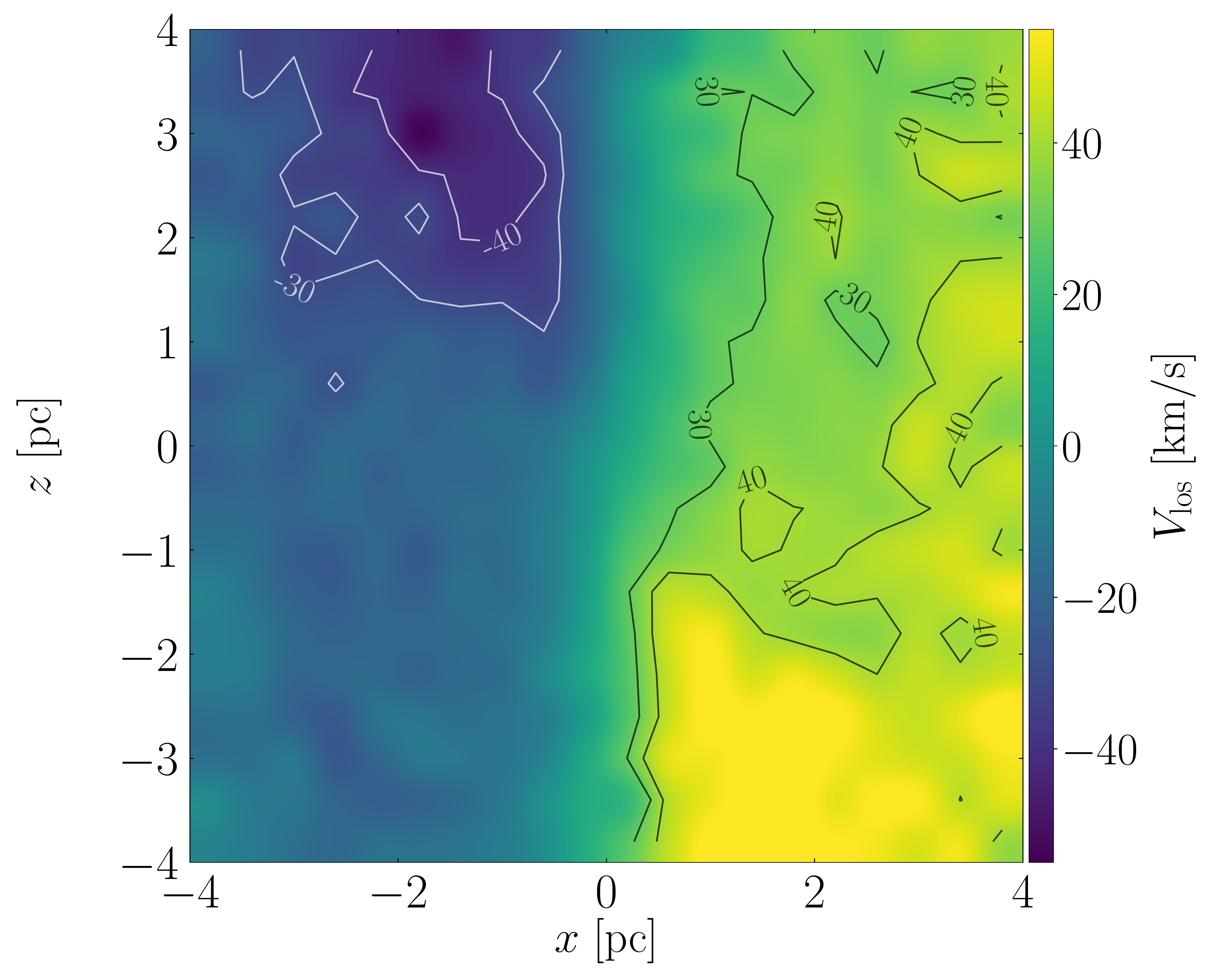}\\
\includegraphics[width=0.45\textwidth]{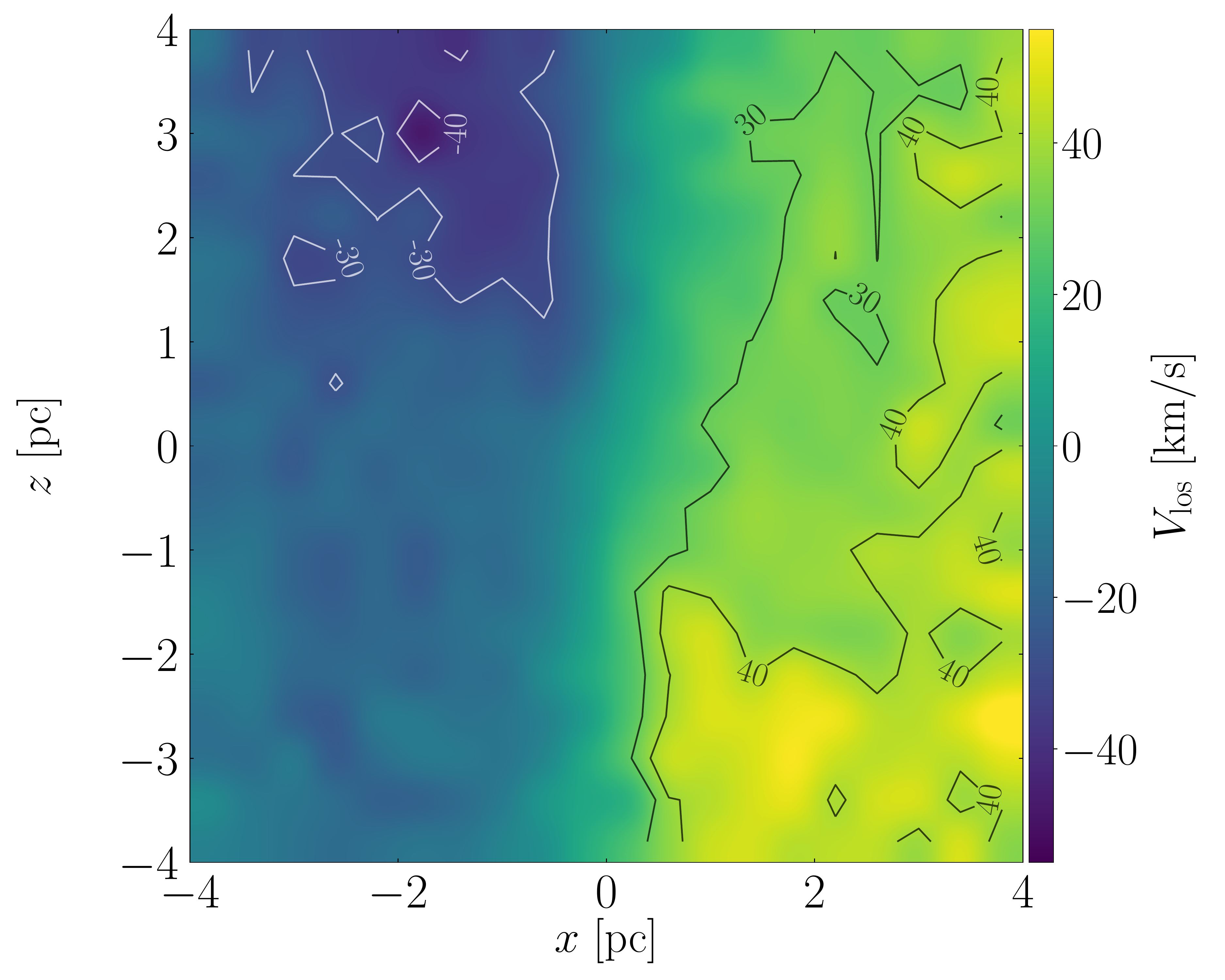}
\includegraphics[width=0.45\textwidth]{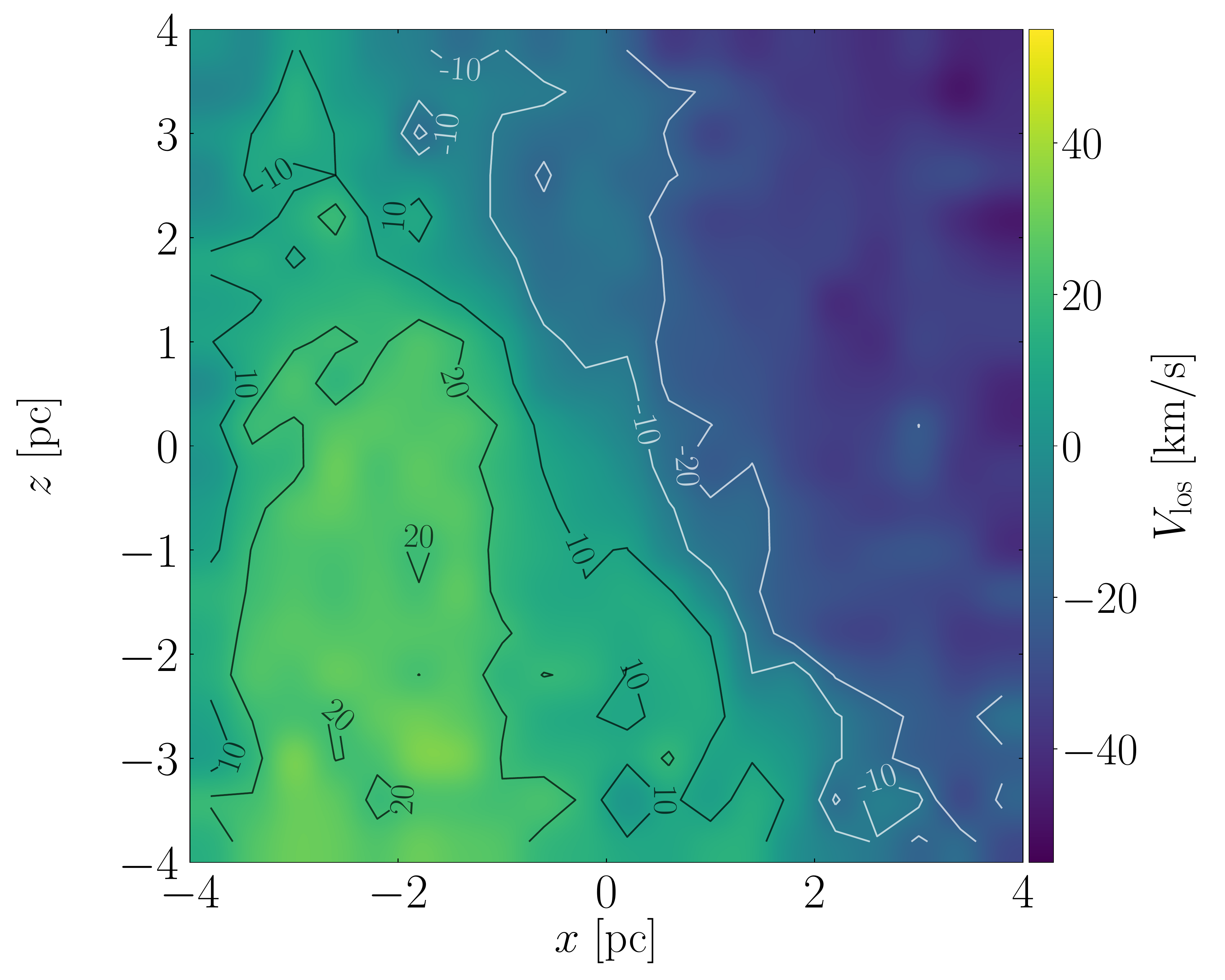}\\
\caption{
Average line-of-sight velocity map in the inner 4 pc for the simulation taken at 200 Myr, assuming i) no NC rotation, ii) NC and former SC members rotate, iii) only NC stars rotate, iv) the NC motion is retrograde compared to the motion of former SC members. Contour lines identify the loci of regions characterised by the same los velocity. As in Figure \ref{fig:los}, the contours are limited to $[-20,-10,10,20]$ km s$^{-1}$ for case with no NC rotation and $[-40,-30,30,40]$ km s$^{-1}$ otherwise.
}
\label{fig:new2}
\end{figure}

Figure \ref{fig:new2} shows the map of the average LOS velocity calculated for the four cases depicted above assuming a total simulation time of 200 Myr.
In all the models explored, the SC debris appears as a diagonal overdensity in the maps. Note that this peculiar distribution is fully driven by the SC debris, as the NC rotation axis in this frame is assumed to be parallel to the Z-axis.

\section{The lifetime of the kinematical signatures}
\label{app:B}

\begin{figure}
\centering
\includegraphics[width=0.5\columnwidth]{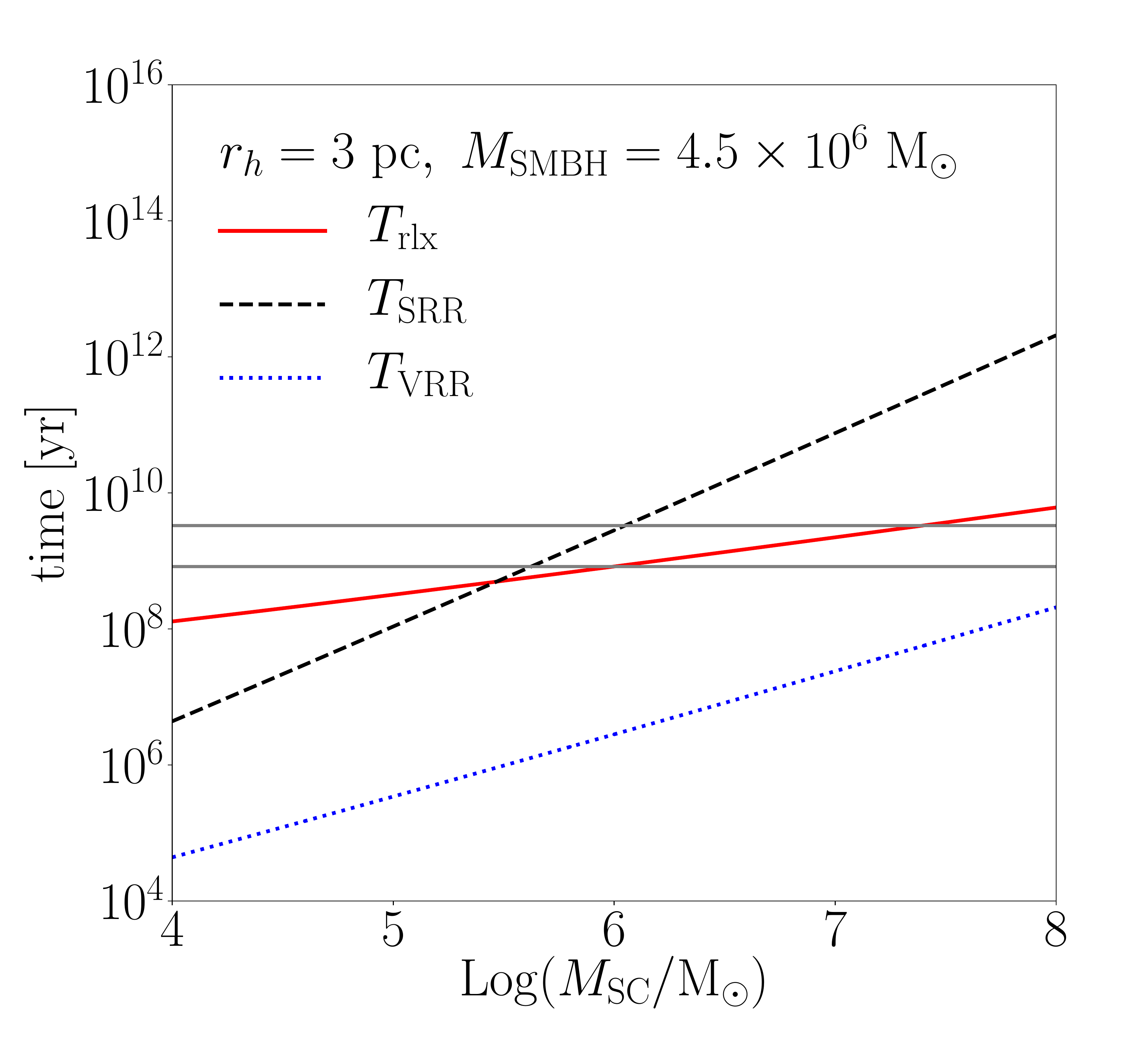}
\caption{Two-body relaxation (straight red line), SRR (dashed black line), and VRR (dotted blue line) timescales as a function of the cluster mass assuming an half-mass radius of 3 pc and a SMBH mass of $4.5\times 10^6\Ms$. Horizontal lines mark the relaxation time for the infalling SC and the whole NC.}
\label{fig:reso}
\end{figure}

In Section \ref{sec:scinfall} we demonstrated that the SC debris remains kinematically distinguishable from the NC for a timescale of at least 100 Myr. Upon dispersal, the SC debris will mix with the NC via {\it two-body relaxation} over a timescale \citep{spitzer71a,bt}
\begin{equation}
T_{\rm rlx} = \frac{6.2 {\rm ~Gyr}}{\ln (0.4M_\nc /\langle m \rangle )}
\left( \frac{M_\nc}{10^7\Ms} \right)^{1/2} 
\left( \frac{1\Ms}{\langle m \rangle} \right)
\left( \frac{r_h}{4{\rm ~pc}} \right)^{3/2}.
\end{equation}
Further mechanisms that can affect the relaxation process are the so-called {\it scalar} (SRR) and {\it vector resonant relaxation} (VRR) \citep{alexander17}. These mechanisms are driven by torques on stellar orbits around the central SMBH \citep{rauch96}. For disc-like configurations, this process can induce warps \citep{kocsis11,perets18} and the formation of spiral arm substructures \citep{perets18}, without affecting the configuration of the system. The timescales for SRR and VRR are \citep{alexander17}
\begin{align*}
T_{\rm SRR} =& \frac{1}{2}\frac{M_\nc}{M_\bh}\ln[M_\bh/\langle m \rangle] T_{\rm rlx}\simeq 40 ~T_{\rm rlx}  , \\
T_{\rm VRR} =& \left(\frac{\langle m \rangle}{M_\nc}\right)^{1/2} T_{\rm SRR} \simeq 8\times 10^{-3} ~T_{\rm rlx},
\end{align*}
for a given mass of the Galactic NC and SMBH. Because the NC's radius of $\sim 4$ pc is larger than the radius of influence of the SMBH,
we only expect a minor influence of the central black hole on the 
relaxation process. Figure \ref{fig:reso} shows how these timescales vary as a function of the cluster mass. The relaxation time for the Galactic NC is $\sim 3$ Gyr. SRR is slower than two-body relaxation, whereas VRR operates on timescales $\sim 10^2$ Myr. Since VRR preserves the structure of the system and $T_{\rm SRR} >> T_{\rm rlx}$, we can assume that VRR and the relaxation times bracket the lifetime of the kinematical features, suggesting that the SC disruption should have happened not more than $0.1-3$ Gyr ago. 

Nonetheless, we note that rotation can lenghten the half-mass relaxation time by a factor up to 3, at least in the simplistic case of a spherically symmetric rotating system  \citep{longaretti96}. However, whether this lengthening of the relaxation time holds for stellar systems harboring an SMBH and a complex substructure (e.g. the SC debris) is unclear.

\section{Semi-analytic modelling of stellar system inspiral}
\label{app:C}

We model the inspiral of the cluster  with a semi-analytic model where dynamical friction operates on a timescale $\tau_\df$ which can be written as \citep{ASCD14a,ASCD15He}
\begin{equation}
\tau_\df = g(e_\gc,\gamma)T_{\rm cr}\left(\frac{M_\gc}{M_g}\right)^{-\alpha}\left(\frac{r_{\rm apo}}{r_g}\right)^{\beta},
\label{eq:df}
\end{equation}
where 
\begin{equation*}
g(e_\gc,\gamma) = (2-\gamma)\left[e_\gc + a_1\left((2-\gamma)^{-a_2} + a_3  \right)(1-e_\gc) \right]
\end{equation*}
is a weak function of the cluster's orbital eccentricity $e_\gc$, $(a_1,~a_2,~a_3) = (2.63,~2.26,~0.9)$, $T_{\rm cr}$ is its orbital period, $r_{\rm apo}$ the SC orbital apocenter, and $\alpha = -0.67$, $\beta = 1.76$, and $(M_g,r_g,\gamma)$ are the host galaxy total mass, length scale and slope of the density profile \citep[see][for further details]{ASCD15He}. Equation~\ref{eq:df} can be used to place constraints on the distance at which the SC formed, which we denote with $r_0$.  Assuming that the decay process can be approximated as an infinite sequence of adiabatic decays, the cluster's infall rate can be written as:
\begin{equation}
    \frac{\derd r}{\derd t} = -\frac{r}{\tau_\df(r)},
\end{equation}
whose integration leads to 
\begin{equation}
    r(t) = r_{\rm apo} \left(1-\frac{\beta t}{\tau_\df(r_{\rm apo})}\right)^{1/\beta},
    \label{eq:rtime0}
\end{equation}
being $r_{\rm apo}$ the initial orbital apocenter. The equation above does not account for the SC's mass loss induced by Galactic tidal forces. Mass loss can be included in the model by discretizing the cluster's orbit in $n$ segments, each covered in a time interval $t_n$, such that the position of the SC at time $t_n$ is given by
\begin{equation}
    r_n \equiv r(t_n) = r_{\rm apo} \displaystyle \prod_{i=1}^{n} \left[1-\frac{\beta t_i}{\tau_\df(M_{i-1}, r_{i-1})}\right]^{1/\beta},
    \label{eq:rtime}
\end{equation}
with ($M_{i-1}, r_{i-1}$) the mass and position of the SC at time $t_{i-1}$. Note that assuming $M_{i} = $ constant implies equality between Equations \ref{eq:rtime0} and \ref{eq:rtime}.
As the SC migrates inward, its size will be limited by the Jacobi radius, i.e. the radius beyond which stars are no longer bound to the SC
\begin{align}
    r_J(t) &\simeq r_n \left(\frac{M_n}{3M_g(r_n)}\right)^{1/3}, \label{eq:rtida}\\
    M_n &\equiv M_n(r_J(t)), \label{eq:mass}
\end{align}
being $M_g(r_n)$ the Galactic mass enclosed within the SC orbit, and $M_n$ the SC mass enclosed within the Jacobi radius.  
Moreover, we assume that the orbit of the infaller circularizes over time 
 due to dynamical friction. For the sake of simplicity, we assume an exponential decline. We postpone a discussion of the role of eccentricity to Appendix \ref{app:C}. Combining Equations \ref{eq:rtime}, \ref{eq:rtida}, and \ref{eq:mass}, we can follow the SC infall and disruption processes once we set the initial values of its mass, orbital radius, and eccentricity.

As discussed in Section  \ref{sec:origin}, we use the equations above to follow the orbital evolution of the SC "progenitor" to a distance of 50 pc from the Galactic center, distinguishing between a Galactic (GAL) and Extragalactic (EXT) origin. In sample GAL, the SC progenitor is modelled as a \cite{Plum} sphere with half-mass radius of $r_{\rm half} = 1-4$ pc and mass assigned in the range $\Log M_\gc = 5-7\Ms$. The cluster's initial apocenter is selected between $r_{\rm apo} = 0.5-50$ kpc, whereas the initial eccentricity is assumed to be $e_\gc = 0.5$, following Gaia DR2 observational constraints \citep{Gaia18}. In sample EXT we assume that the SC progenitor was originally a dwarf galaxy, which is modelled as a \cite{Deh93} sphere with an inner slope of the density profile $\gamma = 0.5$, mass selected in the range $M_\gc = 3.2\times 10^8 - 3.2\times 10^{10}\Ms$ and scale radius in the range $r_c = (100-300)$ pc. The initial apocenter is selected in the range $r_{\rm apo} = 10-300$ kpc and the eccentricity is set to $e_\gc = 0.5$. In both samples, $M_\gc$ and $r_{\rm apo}$ are drawn from log-flat distributions, and we set the progenitor's formation time to 10 Gyr ago. 
In sample GAL, the Galactic bulge is the component that mostly contributes to dynamical friction and tidal forces, while in sample EXT the dark matter halo is the dominant component, modelled as a Dehnen sphere with total mass $M_{\rm DMH} = 10^{12}\Ms$, scale length $r_{\rm DMH} = 30$ kpc, and slope $\gamma = 1$ \citep{dutton14}. We note that both the Galactic bulge and the halo are accounted for to solve Eq.~ \ref{eq:rtime}-\ref{eq:mass}.

\section{The role of the infaller's progenitor orbital eccentricity}
\label{app:D}

In order to evaluate the importance of the orbital eccentricity of the infaller's progenitor, we considered an additional EXT model
assuming $e_\gc = 0.8$, as inferred from cosmological simulations of galactic satellites at high redshift \citep{wetzel11}.

\begin{figure*}
\includegraphics[width=0.48\textwidth]{prova_mappa3}
\includegraphics[width=0.48\textwidth]{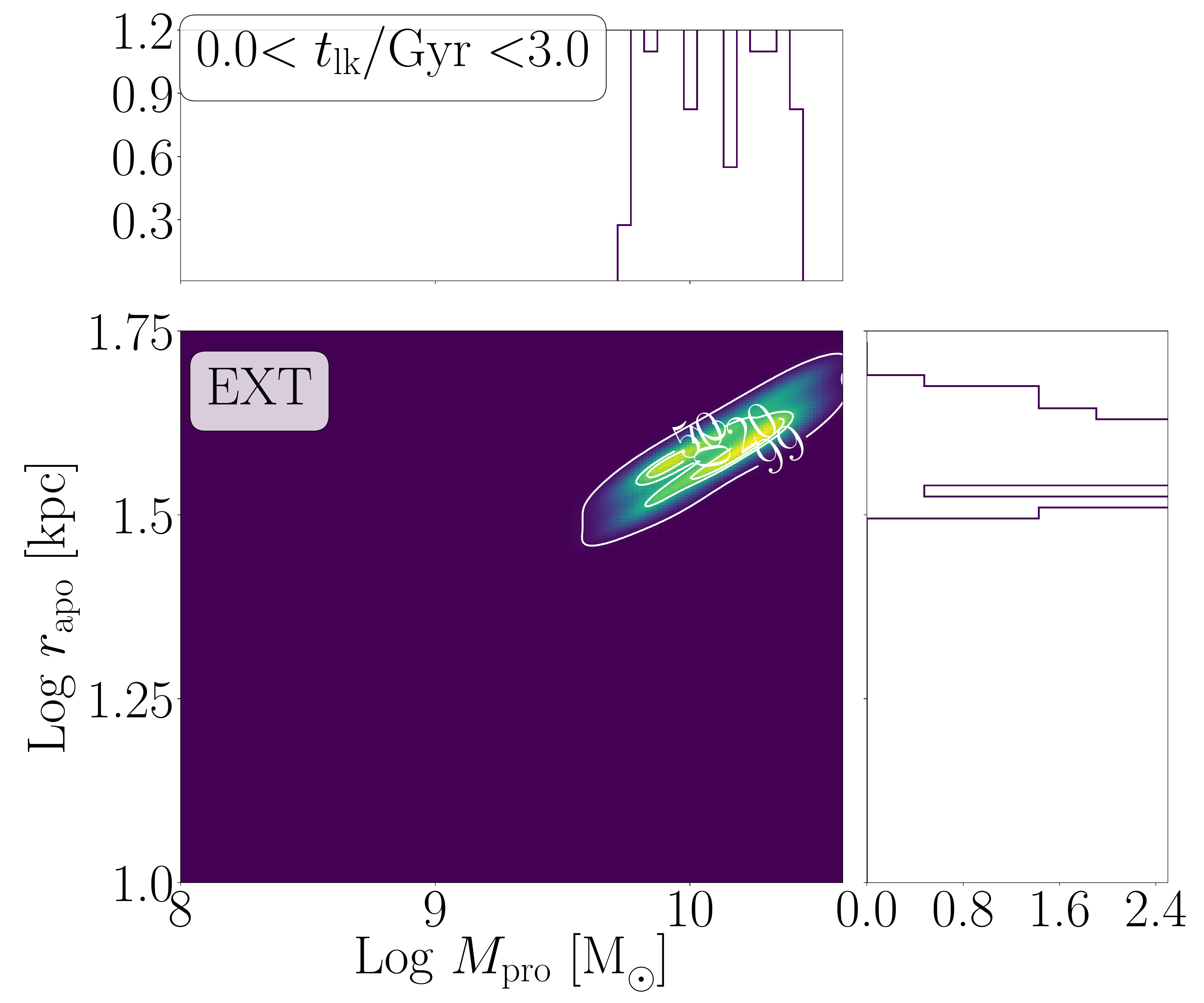}\\
\includegraphics[width=0.48\textwidth]{prova_mappa5} \includegraphics[width=0.48\textwidth]{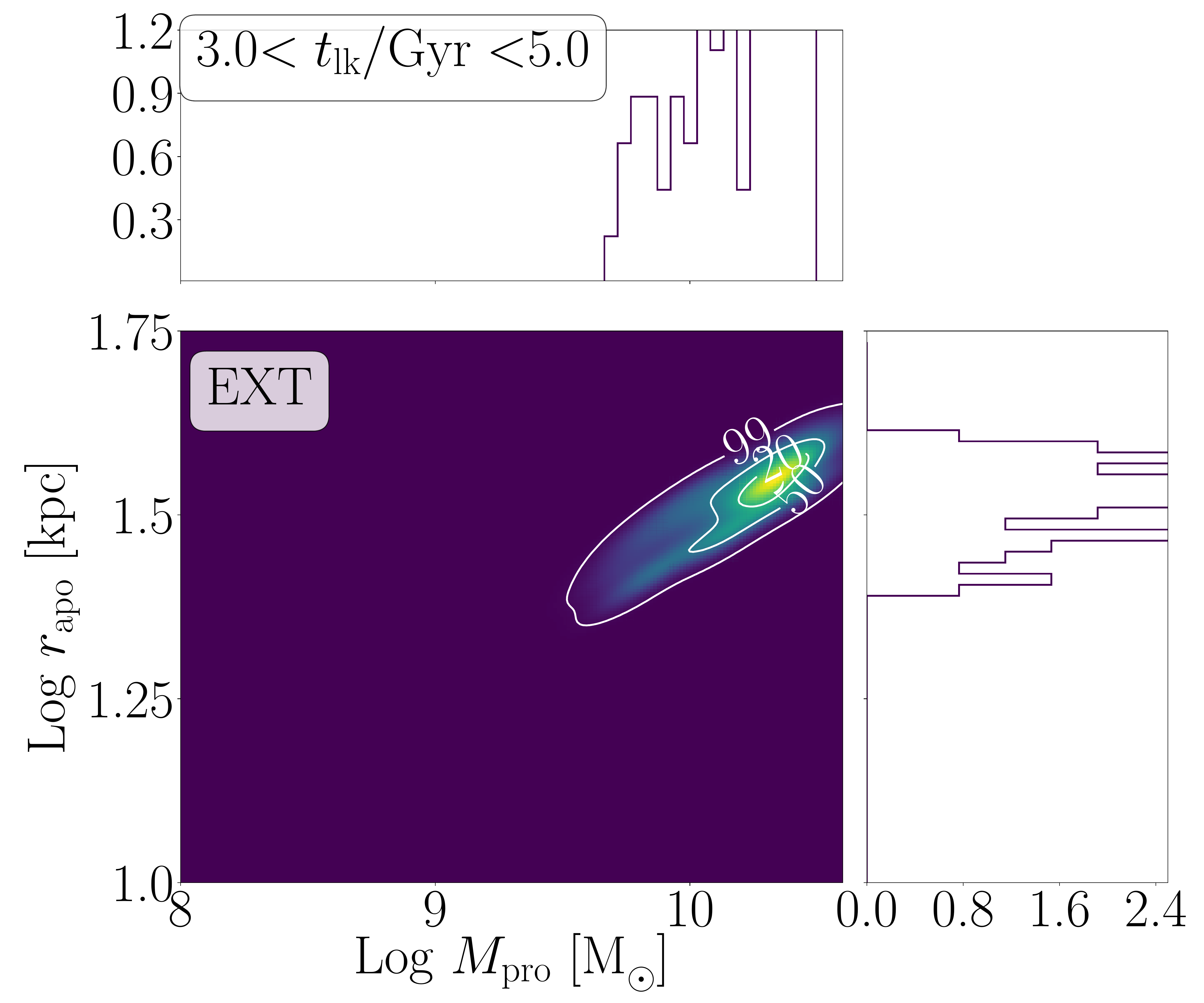}\\ 
\includegraphics[width=0.48\textwidth]{prova_mappa10} \includegraphics[width=0.48\textwidth]{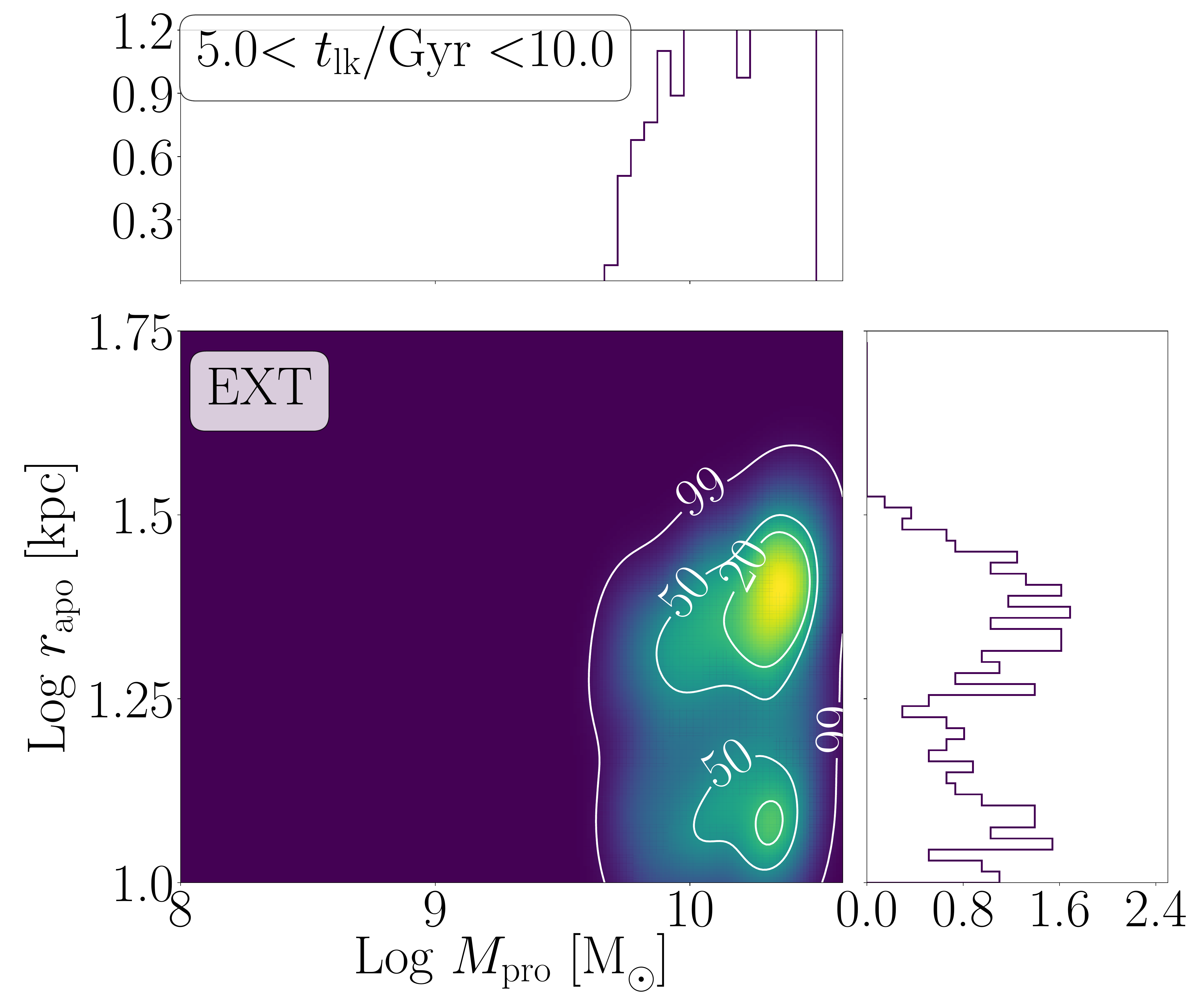} 
\caption{
Left(right) panels: combined surface distribution of initial mass and galactocentric distance in the case of an extragalactic origin, assuming an initial eccentricity $e_{\SC} = 0.5$($0.8$).
}
\label{fig:rori2}
\end{figure*}

A larger eccentricity implies i) a shorter dynamical friction time and therefore a larger possible original distance for the same age ii) a more effective mass loss at pericenter due to a smaller tidal radius for the infaller, leading to larger progenitor initial masses. 
This is highlighted in Figure \ref{fig:rori2}, which shows the combined distributions of initial mass and Galactocentric distance for infallers in the EXT model. We find that a larger value of initial eccentricity would imply a progenitor with mass narrowly peaked around $M_{\SC} \gtrsim 10^{10}\Ms$ and initial position of either $\sim 300$ kpc (late infall) or $\sim 10-100$ kpc (early infall).

\section{A mock stellar catalogue of the Galactic center}
\label{data}

Despite the limitations of our current approach, our model represents the first test-bed to be compared against upcoming proper motions of stars in the Galactic center. In this context, we make the following simulation data publicly available: stellar positions ($x$, $y$, $z$), velocities ($vx$, $vy$, $vz$), X-component of the angular momentum per unit mass ($J_z$, assuming that this is aligned with the rotation axis of the NC), and the ellipticity parameter $J_z/J$ for all stars orbiting inside the innermost 5 pc. A flag is used to identify former SC members (flag=1) and NS stars (flag=0). All quantities refer to times $T = 100$ Myr and $T=200$ Myr and are taken from the post-processed data in which rotation is added to the NC stars only, as detailed in Sections \ref{sec:scinfall} and \ref{app:A}. We caution that this process is not fully self-consistent, and a more realistic exploration of the role of NC rotation will be presented in a future work. Nonetheless, this ``mock'' catalogue of stellar motion inside the central 5 pc of the Galaxy can provide a useful reference when attempting to interpret current and future data.
A small excerpt of the catalogue is given in Table \ref{tab:mock}.

\clearpage
\begin{table*}
\centering
    \caption{Orbital properties of stars orbiting in the inner 5 pc of the Galaxy. Col. 1-3: positions in pc. Col. 4-6: velocities in km/s. Col. 7-8: component of the angular momentum per unit mass parallel to the NC rotation axis, in units of m$^2$/s, and normalized to the total angular momentum.  Col. 9: flag identifying former SC members (F=1) or NC stars (F=0). Data refer to a simulated time of $100$ Myr (top) and $200$ Myr (bottom). Full tables are available online. Here, data are truncated to the second decimal point.}
    \label{tab:mock}
    \begin{tabular}{ccccccccc}
    \hline
    \hline
    $x$ & $y$ & $z$ & $v_x$ & $v_y$ & $v_z$ & $J_z$ & $J_z/J$ & F\\
    pc & pc & pc & km/s & km/s & km/s & m$^{2}$/s &  & \\
    \hline
    \multicolumn{9}{c}{$T = 100$ Myr}\\
    \hline
    $-3.73\times 10^{-1}$ & $-7.28\times 10^{-1}$ & $-1.35$ & $5.93 $ & $7.37\times 10^{1} $ & $-9.20\times 10^{1}$ & $ -7.14\times 10^{20}$ & $ -1.33\times 10^{-1}$ & $ 1 $\\
    $9.31\times 10^{-2} $ & $1.80 $ & $-2.47\times 10^{-1}$ & $-9.01\times 10^{1}$ & $3.32\times 10^{1}$  & $ 1.22\times 10^{2}$ & $ 5.10\times 10^{21} $ & $5.85\times 10^{-1} $ & $1 $\\
    $2.34 $ & $1.00 $ & $-3.56$ & $-7.18\times 10^{1}$ & $5.06\times 10^{1}$  & $ 3.27\times 10^{1}$ & $ 5.87\times 10^{21} $ & $5.65\times 10^{-1} $ & $1 $\\
    $-1.74\times 10^{-1}$ & $-7.24\times 10^{-1}$ & $-1.60\times 10^{-1}$ & $5.50\times 10^{1} $ & $1.38\times 10^{2} $ & $-1.50\times 10^{2}$ & $ 4.86\times 10^{20} $ & $1.15\times 10^{-1} $ & $1 $\\
    $3.01\times 10^{-2} $ & $3.28 $ & $2.54\times 10^{-1} $ & $-4.07\times 10^{1}$ & $-2.19\times 10^{1}$ & $ 1.11\times 10^{2}$ & $ 4.10\times 10^{21} $ & $3.36\times 10^{-1} $ & $1 $\\
    \multicolumn{9}{c}{...}\\
    \multicolumn{9}{c}{...}\\
    \multicolumn{9}{c}{...}\\
    \hline
    \multicolumn{9}{c}{$T = 200$ Myr}\\
    \hline
    $3.41\times 10^{-1} $ & $ 9.23\times 10^{-1}$ & $ -7.75\times 10^{-1}$ & $ -9.51\times 10^{1}$ & $ 9.76\times 10^{1} $ & $ 5.32\times 10^{1}  $ & $3.73\times 10^{21} $ & $ 6.63\times 10^{-1} $ & $ 1$ \\ 
    $7.69\times 10^{-1} $ & $-1.07$ & $ -2.42$ & $ -1.24\times 10^{1}$ & $ 1.02\times 10^{2} $ & $ -5.30\times 10^{1} $ & $2.00\times 10^{21} $ & $ 2.03\times 10^{-1 }$ & $ 1$ \\ 
    $8.18\times 10^{-1} $ & $-9.15\times 10^{-1}$ & $ 1.60 $ & $ 3.44\times 10^{1} $ & $ -1.64\times 10^{2}$ & $ -4.00\times 10^{1} $ & $-3.17\times 10^{21}$ & $ -3.11\times 10^{-1}$ & $ 1$ \\ 
    $-2.78\times 10^{-1}$ & $ 2.12\times 10^{-1}$ & $ 5.02\times 10^{-1} $ & $ -4.00\times 10^{1}$ & $ -6.27\times 10^{1}$ & $ 1.41\times 10^{2}  $ & $8.01\times 10^{20} $ & $ 3.73\times 10^{-1} $ & $ 1$ \\ 
    $3.55\times 10^{-2} $ & $-3.28$ & $ 1.08 $ & $ 2.29\times 10^{1} $ & $ 1.48\times 10^{-1} $ & $ -1.22\times 10^{2} $ & $2.32\times 10^{21} $ & $ 1.83\times 10^{-1} $ & $ 1$ \\ 
    \multicolumn{9}{c}{...}\\
    \multicolumn{9}{c}{...}\\
    \multicolumn{9}{c}{...}\\
    \hline
    \end{tabular}
\end{table*}

\clearpage

\footnotesize{
\bibliographystyle{mn2e}
\bibliography{ASetal2015}
}

\end{document}